\documentclass[review]{elsarticle}

\usepackage{graphicx, lineno, hyperref, amsmath, float, makecell}
\modulolinenumbers[5]

\journal{Journal of Empirical Finance}

\begin{document}

\begin{frontmatter}

\title{A New Multivariate Predictive Model for Stock Returns}

\author[mymainaddress]{Jianying Xie\fnref{myfootnote} \corref{mycorrespondingauthor}}
\fntext[myfootnote]{Supervised by Professor Jiti Gao, Department of Econometrics and Business Statistics, Monash University, Melbourne, 3800, Australia}
\cortext[mycorrespondingauthor]{Corresponding author}
\ead{jxie73@student.monash.edu}


\address[mymainaddress]{Department of Econometrics and Business Statistics, Monash University, Melbourne, 3800, Australia}

\begin{abstract}
	One of the most important studies in finance is to find out whether stock returns could be predicted. This research aims to create a new multivariate model, which includes dividend yield, earnings-to-price ratio, book-to-market ratio as well as consumption-wealth ratio as explanatory variables, for future stock returns predictions. The new multivariate model will be assessed for its forecasting performance using empirical analysis. The empirical analysis is performed on S\&P500 quarterly data from Quarter 1, 1952 to Quarter 4, 2019 as well as S\&P500 monthly data from Month 12, 1920 to Month 12, 2019. Results have shown this new multivariate model has predictability for future stock returns. When compared to other benchmark models, the new multivariate model performs the best in terms of the Root Mean Squared Error (RMSE) most of the time.
\end{abstract}

\begin{keyword} 
	\texttt{ financial time series, stock return prediction, multivariate model, level of nonstationarity, co-integration, multi-step ahead forecasting}
\end{keyword}

\end{frontmatter}

\section{Introduction}
	One of the most important studies in finance is to find out whether stock returns can be predicted. This is because, movements in the stock market, for example, an increase or decrease in stock prices that lead to any changes in stock returns, can result in a remarkable economic impact on the entire economy and businesses as well as the individual consumer. There have been debates over the past few decades on whether stock returns can be predicted. Some people support market efficient theory and argue that stock returns are not predictable while others argue that stock returns are predictable due to the rational variation contained in the stock returns. Many studies in this community have concluded that stock returns are predictable \cite{fama1988permanent, fama1990stock, campbell2000asset, lettau2001consumption, campbell2006efficient, lettau2008reconciling}. Those studies suggested that stock returns’ predictability is from the rational variation in expected returns, where the expected returns contain a time-varying component that implies predictability of future returns. Lettau and Ludvigson (2001) \cite{lettau2001consumption} demonstrate that it is now widely accepted that excess returns are predictable by variables such as dividend-price ratios, earnings-to-price ratios, dividend-to-earnings ratios and an assortment of other financial ratios. However, according to Welch and Goyal (2008) \cite{welch2008comprehensive}, profession has yet to find some variable that has meaningful and robust empirical equity premium forecasting power both in-sample and out-of-sample, whether the stock returns are predictable remains unsettled .
	
	To predict stock returns, most studies use financial ratios, such as dividend yield, earnings-to-price ratio, and book-to-market ratio as predictors \cite{fama1989business, fama1990stock, lamont1998earnings, stambaugh1999predictive, welch2008comprehensive}. There is evidence supporting that these financial ratios could forecast stock returns. Rozeff (1984) \cite{rozeff1984dividend} justify the dividend yield is directly related to the prediction for future stock returns. Higher dividend yield indicates higher stock returns. According to Lamont (1998) \cite{lamont1998earnings}, higher earnings forecast low returns. This is because investors would require high expected returns in recessions, and relatively lower expected returns in economic booms. Since earnings vary with the economic activities, current earnings can predict future stock returns. That is, the correlation between earnings and business as well as the economic conditions gives the predictive power for earnings for stock returns prediction. Since the publication of Fama (1992) \cite{eugene1992cross}, the book-to-mark ratio has become a popularly used ratio for stock returns prediction. According to Kothari and Shanken (1997) and Pontiff and Schall (1997) \cite{kothari1997book, pontiff1998book}, their empirical analysis has shown that the book-to-market ratio helps predict stock returns. In general, it is found that financial ratios, including dividend yield, earnings-to-price ratio, and book-to-market ratio, have predictability for future stock returns. Thus, in this research, dividend yield, earnings-to-price ratio, and book-to-market ratio are included as predictors. Whether these predictors have predictive power for future stock returns has also be investigated.

	If stock returns can be predicted, it provides a gauge of risk for the companies and investors. Based on the stock returns prediction, investors can gain a significant amount of money from the stock market, and the loss from the investment in the stock market can be controlled. Hence, stock return predictions affect resource allocation in the economy. The multivariate model combines different regressors, including dividend yield, earnings-to-price ratio, book-to-market ratio as well as consumption-wealth ratio for stock returns prediction. The multivariate model’s predictability compares to other benchmark models including the popular historical mean model and autoregressive model. Additionally, this paper aims to find out whether the financial ratios, including dividend yield, earnings-to-price ratio, and book-to-market ratio, as well as consumption-wealth ratio have predictability for future stock returns. The new multivariate model is expected to take into account multiple regressors and to have short, intermediate, and long-horizon forecasting power. In this research, whether the level of nonstationarity due to the financial ratios could be reduced, resulting in improving predictability for future stock returns, have also be explored.
	
	However, there are some issues in stock returns prediction. Stambaugh pointed out that the regression disturbance could be correlated with regerssors’ innovation when the stock returns are predicted by lagged stochastic regressors, such as dividend yield \cite{stambaugh1999predictive}. In this case, the Ordinary Least Squared (OLS) estimator with finite-sample properties would deviate from those in the standard regression setting. Many researchers have found that many predictive variables used for stock returns prediction are unit root or nearly unit root processes \cite{nelson1993predictable, stambaugh1999predictive}. In this paper, OLS is the chosen method for parameters estimation since the model is constructed to be linear in parameters. Then, the in-sample test is conducted to study the estimated coefficients’ in-sample predictability. The out-of-sample prediction utilise the recursive window for forecasting. This enables us to compare different models and investigate their out-of-sample forecasting performance, which is our main focus in this research. The error measure Root Mean Squared Error (RMSE) is used as a gauge for the out- of-sample forecasting performance. The new multivariate model results in having strong predictability for future stock returns, contributing to current studies in stock returns prediction. In the meantime, the variables are considered to be persistent, which will lead to biased estimated coefficients and misleading inference on predictors’ predictive ability. In particular, Lettau and Nieuwerburgh (2008) \cite{lettau2008reconciling} justify that financial ratios are extremely persistent even though they are proven to have predictive ability for stock returns. In this case, the forecasting relationship of returns and financial ratios exhibits significant instability over time. The persistency adds difficulty in stock returns prediction. Since the financial ratios are considered to be extremely persistent, Goyal and Welch (2003, 2008) \cite{goyal2003predicting, welch2008comprehensive} suggested that dividend yields primarily predict themselves over a short horizon. It is considered that the financial ratios only have predictability at the long horizon. This paper reduce the impact of persistency due to the financial ratios as well as involve their long-horizon forecasting power into the model.
	
	Besides financial ratios, there is also research using the consumption-wealth ratio as a predictor for future stock returns prediction. The consumption-wealth ratio is calculated by the deviations from the common trend in consumption, asset holdings, and labor income for predicting stock market fluctuations. Lettau and Ludvigson (2001) \cite{lettau2001consumption} found that the consumption-wealth ratio is a strong predictor for future stock returns. The research made this conclusion based on the evidence that the log consumption-aggregate wealth (human capital plus asset holdings) ratio summarizes expected returns on aggregate wealth, or the market portfolio. Since human capital is not observable, observable variables including consumption, asset holdings, and labor income are used instead. It is considered that the consumption-wealth ratio is a better predictor than dividend yield, dividend payout ratio as well as other popular financial ratio predictors at short and intermediate horizons \cite{lettau2001consumption, kostakis2015robust}. It is able to capture information about future stock returns while other popular financial ratios cannot. Additionally, evidence has shown consumption-wealth ratio can improve out-of-sample forecasting performance to a large variety of models, using postwar data. Thus, in this research, the consumption-wealth ratio is also included as one of the predictors for forecasting stock returns. Whether the consumption-wealth ratio will increase predictive power in the multivariate model has been examined in this research.
	
	Furthermore, many studies on stock returns prediction are conducted for univariate regressor. They tested its predictability individually, such as Stambaugh (1999) and Welch and Goyal (2008) \cite{stambaugh1999predictive, welch2008comprehensive}. Ang and Bekaert (2007) \cite{ang2007stock} demonstrate that a univariate regression model might not be sufficient to capture all the predictable components in returns when they only have a univariate dividend yield regression, as it provides a relatively bad proxy for stock returns. However, when it comes to using short rate and dividend yield together, the regression is improved particularly at the short horizon. When they conduct the empirical analysis, it is shown that the coefficient on the dividend yield is larger in the bivariate model than in the univariate model. That is, in the univariate model, the marginal effect of dividend yield is understated. The univariate model might suffer from the omitted variable bias. The predictability of dividend yield in a univariate regression is underestimated. Thus, there are risks to regress stock returns on a single variable which will provide a biased estimation. It is also recommended to use the entire information set to assess the predictability rather than isolating each variable (Kostakis et al, 2015) \cite{kostakis2015robust}. Hence, a multivariate model is used in this research, which combines dividend yield, earnings-to-price ratio, book-to-market ratio, as well as consumption-wealth ratio for stock returns prediction. Based on the existing studies, a new multivariate model is constructed. It is expected that this multivariate model can help with solving some problems in the existing literature. In the following section, the new multivariate model will be defined in detail.

	The key findings are as follows. Firstly, the new multivariate model constructed in this research does have predictability for future stock returns. Secondly, this new multivariate model is able to beat the historical mean model and autoregressive model when conducting multi-step ahead forecasting. Thirdly, when dividend yield, earnings-to-price ratio, book-to- market ratio, as well as consumption wealth ratio are included in a model collectively to predict future stock returns, they help with improving the predictive power for out-of-sample forecasting. Last but not least, the level of nonstationarity due to the persistent predictors can be reduced by introducing an exponential term. The exponential term will be discussed in the following section and the key findings will be further discussed in Section 5.
	
	The structure of the paper is as follows: Section 2 provides the methodology for stock returns prediction, including model specification, and econometric methods used for out-of-sample forecasting. Section 3 describes the data source and its description. Section 4 presents the results of the empirical analysis. Section 5 presents the conclusion and findings.

\section{Methodology}
	This section discusses the general form of the new multivariate model which is constructed for stock returns prediction in this research. Since the predictors chosen to perform the stock returns prediction are often considered to be nonstationary, the unit root test will be conducted to investigate the statistical significance of the nonstationarity. The Ordinary Least Squared Estimation (OLS) method will be used for the estimation of the coefficients. The F- test and t-test will be conducted to investigate the in-sample explanatory variables’ predictability. A decision can then be made on which variable should be included or dropped in the out-of-sample prediction based on the F-test and t-test results. With respect to the out- of-sample evaluation, an estimation window method needs to be chosen for out-of-sample forecasting. Then the error measure Root Mean Squared Error (RMSE) can be used as a gauge for the out-of-sample forecasting performance. In this section, the methodology which are used in this research will be discussed.
	
	\subsection{Model Specification}
	    Cai and Gao (2017) \cite{cai2017simple} demonstrate an exponential term $(e^{-\frac{x_t^2}{2}})$ in their nonlinear model for stock returns prediction. This term is introduced to reduce the level of nonstationarity of the predictors and to have a better balance for the regression than a simple linear model. A brief mathematical proof for how the level of nonstationarity is being reduced is included in Appendix 1. A model which combines dividend yield, earnings-to-price ratio, book-to-market ratio, and consumption-wealth ratio for stock returns prediction to take into account all possible available information is desirable. Hence, a multivariate model is used. The new multivariate model is built upon linear parameters, nonlinear predictors, and the exponential term from Cai and Gao (2017) \cite{cai2017simple}. Thus, our new multivariate model of the general form is as equation \ref{multivariatemodelform}.
		\begin{equation}
        \label{multivariatemodelform}
            y_t = \sum \limits _{j=1}^p \theta_j y_{t-j} + \sum \limits _{k=1}^q (\alpha_k + \beta_k x_{t-1, k})e^{-\frac{x_{t-1, k}^2}{2}} + e_t
		\end{equation}
	
		\subsubsection{Predictors Selection}
			Firstly, lagged stock return $(y_{t-j})$ is chosen as one of the predictors. It is considered that the predictability of stock returns is the result of the rational variation in stock returns across time. $y_{t-j}$ is used to capture past information, such as prior time-varying risk premium and the time-varying components in previous stock return, which might have some predictability for future stock returns according to Fama and French (1989) \cite{fama1989business} and Lettau and Nieuwerburgh (2008) \cite{lettau2008reconciling}.
		
			Secondly, the dividend yield, earnings-to-price ratio, and book-to-market ratio are chosen to be the $x_{t-1, k}$ predictors in equation \ref{multivariatemodelform}. The $(t - 1)$ indicates one time period lagged for the $x$ predictors.
		
			For the dividend yield, it is a widely used predicting variable in stock returns prediction in lots of studies such as Fama and French (1989), Fama (1990), Pesaran and Timmermann (1995), and Lamont (1998) \cite{fama1989business, fama1990stock, pesaran1995predictability, lamont1998earnings}. There is evidence supporting that dividend yield retains good statistical significance in the in-sample estimation and could forecast stock return (Goyal et al., 2003) \cite{goyal2003predicting}. For the earnings-to-price ratio, it is considered that it can reflect business condition, which will contribute to stock returns prediction (Fama and French, 1989; Fama, 1990; Pesaran et al, 1995) \cite{fama1989business, fama1990stock, pesaran1995predictability}. For the book-to-market ratio, it is another commonly used ratio in lots of studies for stock returns prediction. Many researchers, such as Fama and French (1988), Kothari and Shanken (1997), and Pontiff and Schall (1998) \cite{fama1988permanent, kothari1997book, pontiff1998book}, show that the book-to-market ratio helps predict stock returns. These popular financial ratios are used as forecasting variables in this multivariate model. Their predictability for future stock returns will be also studied in this research. It is expected that these variables could help with increasing the predictability of the multivariate model. The results can then be compared with existing studies.
		
			Lastly, the consumption-wealth ratio, which gives the common trend deviation in the common trend of consumption, asset holdings, and labor income (Lettau et al., 2001) \cite{lettau2001consumption}, is included as one of the predictors. This is because the consumption-wealth ratio is considered as a good proxy for market expectations of future asset returns with short and intermediate forecasting power (Lettau et al., 2001, Kostakis et al., 2015) \cite{lettau2001consumption, kostakis2015robust}. A multivariate model which has predictability for future stock returns at the short and intermediate horizon is desirable.
		
			Since the consumption-wealth ratio is generally accepted as a stationary variable (Lettau et al., 2001; Welch et al., 2008) \cite{lettau2001consumption, welch2008comprehensive}, and when the consumption-wealth ratio is included as a predictor $x_{t-1, cay}$, equation \ref{multivariatemodelform} needs to be adjusted as:
			\begin{equation}
        	\label{multivariatemodelformadjusted}
        	    y_t = \sum \limits _{j=1}^p \theta_j y_{t-j} + \beta_{cay} x_{t-1, cay} + \sum \limits _{k=1}^q (\alpha_k + \beta_k x_{t-1, k})e^{-\frac{x_{t-1, k}^2}{2}} + e_t
			\end{equation}
		
		\subsubsection{Expected Contributions}
			Since the financial ratios are found to be persistent, which implies unit root, they are nonstationary variables on the right-hand side (RHS) of equation \ref{multivariatemodelformadjusted}. However, stock returns are usually considered to be stationary or at least far less persistent than the predictive variables. In this case, regression’s left-hand side (LHS, stationary stock returns) and RHS (nonstationary, financial ratios) are imbalanced. Therefore, an exponential term $(e^{-\frac{x_{t-1, k}^2}{2}})$ is introduced to reduce the level of nonstationarity on the RHS of equation \ref{multivariatemodelformadjusted}. When the level of nonstationarity is reduced on RHS and stock return is stationary on the LHS, the multivariate model is considered to be balanced. Hence, the predictability of the multivariate model for stock returns is expected to increase.
		
			Moreover, for nonstationary time series, there exists a linear combination with this time series that could help with reducing the level of nonstationarity (Engle and Granger., 1987) \cite{engle1987co}. This is the concept of co-integration. Lee (1996) \cite{lee1996comovements} has investigated that the earnings and dividends are co-integrated using data from S\&P500 during the period 1871 to 1992. The relationship or the restrictions of co-integration will highly affect the forecasting power (Duy and Thoma., 1998) \cite{duy1998modeling}. Since dividend yield and earnings-to-price ratio are predictors in this research, the co-integration relationship should be considered in this new multivariate model. If the co-integration is taken into account in this research, the level of nonstationarity can be reduced and better forecasting performance will be provided.
		
			As mentioned in Section 1, financial ratios are found to have poor short-horizon predictability and often give the long-horizon predictability due to their persistence. Hence, the consumption-wealth ratio is added to take into account the short and intermediate horizon predictability. Also, a combination of multiple regressors that contain financial ratios, macroeconomic ratio (the consumption-wealth ratio), as well as the lagged stock returns which capture past information, could reflect different material at a time. Overall, it is expected that this new multivariate model could take into account multiple regressors and provide predictive power at short, intermediate, and long horizons.
			
		\subsection{Data Examination: Unit Root Test}
		    As mention in Section 1, financial ratios are usually considered to be nonstationary. In this case, the unit root test needs to be conducted to check the statistical significance of the level of nonstationarity. There are many unit root tests including Dickey-Fuller (DF), Augmented Dickey-Fuller (ADF), Pillips-Perron (PP), and KPSS Tests. Amongst these unit root tests, ADF Test will be used in this research. ADF Tests are under the restriction for independent, identically distribution $(i. i. d.)$ or autocorrelated time series. Theoretically, the unit root tests will give the same conclusion on the stationarity for the variables.
		
		    The general procedure of implementing the ADF Test is as follows according to Lütkepohl, Krätzig and Markus (2004):
	    	\begin{itemize}
      			\item Consider a model of the form:
    				\begin{equation}
    				\label{xequation}
    					x_t = \rho x_{t-1e_t} 
    			    \end{equation}
    			    where $x_t$ can be considered to be any one of the time series including dividend yield, earnings-to-price ratio, book-to-market ratio.
				
			    \item Subtracting $x_t$ from both sides of equation \ref{xequation}, it is able to yield:
			
				    $x_t - x_{t-1} = \rho x_{t-1} - x_{t-1} + \mu_t = (\rho - 1) x_{t-1} + \mu_t$
				
				    That is, 
				    \begin{equation}
    				\label{improvexequation}
    					\Delta x_t = \gamma x_{t-1} + \mu_t
    				\end{equation}
    				where $\Delta x_t = x_t - x_{t-1}$ and $\gamma = \rho - 1$.
				
			    \item The ADF Test has the null hypothesis of nonstationarity against an alternative hypothesis of stationarity:
    				\begin{center}
    					$H_0: \gamma = 0 \Leftrightarrow \rho = 1$
    					
    					$H_1: \gamma < 0 \Leftrightarrow |\rho| < 1$
    				\end{center}
				
			    \item Then the test statistic can be calculated by:
    				\begin{equation}
    				\label{teststatistic}
    					t(\hat \gamma) = \frac{\hat \gamma}{se(\hat \gamma)}
    				\end{equation}
    				where $\hat \gamma$ is the ordinary least squares (OLS) estimator of $\gamma$ and 
    				\begin{center}
    					$se(\hat \gamma) = \sqrt{\frac{\sum  _{t=1}^T (\Delta x_t - \hat \gamma x_{t-1})^2}{T \sum _{t=1}^T x_{t-1}^2}}$
    				\end{center}
				
			    \item Decision rule: If $|t(\hat \gamma)| > \textit{Dickey - Fuller t critical value}$, the null hypothesis can be rejected at a predetermined level of significance. There is sufficient evidence to conclude that the times series is stationary. Otherwise, the null hypothesis will not be rejected and the time series is considered to be nonstationary.      			
    	    \end{itemize}
    	    
        \subsection{Ordinary Least Squared (OLS) Estimation Method}
	
    		The OLS estimation method is used in this study to estimate the coefficients since the multivariate model is linear in parameters. The main idea of using OLS to find the estimators is to minimize the Sum of Squared Error (SSR). That is, the coefficients need to satisfy the following equation \ref{ssr}:
    		\begin{equation}
    		\label{ssr}
    			SSR = Min \sum \limits _{t=2}^T(y_t - \sum \limits _{j=1}^p \theta_j y_{t-j} - \sum \limits _{k=1}^q (\alpha_k + \beta_k x_{t-1, k}) e^{- \frac{x_{t-1,k}^2}{2}})^2
    		\end{equation}
    		
    		Let $\theta$ be a matrix where $\theta = \begin{bmatrix} \theta_1 \\ \vdots \\ \theta_p \end{bmatrix}$, $y_j = \begin{bmatrix} y_{t-1} \\ \vdots \\ y_{t-p} \end{bmatrix}$, and 
    $Y = \begin{bmatrix} y_1 \\ \vdots \\ y_t \end{bmatrix}$:
    		\\
    		
    		\begin{equation}
    		\label{thetaequation}
    			(\alpha_k + \beta_k x_{t-1, k}) e^{- \frac{x_{t-1, k}^2}{2}} = \alpha_k e^{- \frac{x_{t-1, k}^2}{2}} + \beta_k x_{t-1, k} e^{- \frac{x_{t-1, k}^2}{2}}
    		\end{equation}
    		
    		Let
    		\begin{equation}
    		\label{muequation}
    			\mu_{t-1, k} = e^{- \frac{x_{t-1, k}^2}{2}}
    		\end{equation}
    		
    		and
    		\begin{equation}
    		\label{vequation}
    			\nu_{t-1, k} = x_{t-1, k} e^{- \frac{x_{t-1, k}^2}{2}}
    		\end{equation}
    		\\
    		
    		and $\gamma_k = \begin{bmatrix} \alpha_k , \beta_k \end{bmatrix}$, $\gamma = \begin{bmatrix} \gamma_1 \\ \vdots \\ \gamma_p \end{bmatrix}$, $\omega_t = \begin{bmatrix} \omega_{t1} \\ \vdots \\ \omega_{tp} \end{bmatrix}$, $\omega_{tk} = \begin{bmatrix} \mu_{t-1, k},  \nu_{t-1, k} \end{bmatrix}$, then the equation \ref{ssr} becomes:
    		\begin{equation}
    		\label{ssrimproved}
    			SSR = Min \sum \limits _{t=2}^T (y_t - \theta^{'} y_j - \gamma^{' \omega_t})^2
    		\end{equation}
    		\\
    		
    		Let $\xi = \begin{bmatrix} \theta \\ \gamma \end{bmatrix}$ and $\varphi =  \begin{bmatrix} y_j \\ \omega_t \end{bmatrix}$, then the OLS estimators for $\xi$ is presented as the following:
    		\begin{equation}
    		\label{xiequation}
    			\hat \xi = \begin{bmatrix} \hat \theta \\ \hat \gamma \end{bmatrix} = \begin{bmatrix} \varphi^{'} \varphi \end{bmatrix}^{-1} \varphi^{'} Y
    		\end{equation}
		
	\subsection{In-sample Predictability Tests}
		According to Inoue and Kilian (2004) \cite{inoue2004bagging}, when the number of predictors is moderately large relative to the sample size, a testing procedure can be implemented to decide which predictors should be included in the out-of-sample forecasting model and which should be dropped. In this subsection, the F-test and t-test will be utilised for testing the in-sample predictability for the explanatory variables.
		
		\subsubsection{The F-test}
			According to equation \ref{multivariatemodelform}, the coefficients which need to be tested in this research are the $\theta_j$ and $\beta_k$. If any one of the estimated coefficients is not statistically significant, theoretically, this explanatory variable should be dropped out from the estimation. The statistical significance is regarded as an implication of predictability. The F-test is conducted to investigate the explanatory variables’ overall predictability. The general procedure of implementing the F-test to investigate whether the explanatory variable is statistically significant is as follows according to Wooldridge (2020) \cite{wooldridge2015introductory}:
			\begin{itemize}
				\item The F-test has the null hypothesis of the coefficient of explanatory variables should all be equal to zero against an alternative hypothesis of at least one of the coefficients should not equal to zero:
					\begin{center}
						$H_0: \theta_j = \beta_k = 0$
						
						$H_1: $ At least one of $\theta_j$ and $\beta_k$ does not equal to zero
					\end{center}
					
				\item The test statistic can be calculated as follows:
					\begin{equation}
					\label{fstats}
						F_{stats} = \frac{(SSR_r - SST_{ur})/q}{SST_{ur} / (n-k-1)} \sim F_{q, {n-k-1}} \ under \ H_0
					\end{equation}
					where $q$ is the number of restrictions, and $F_{q, {n-k-1}}$ is an $F$ distribution with $(q, n - k - 1)$ degrees of freedom. $q$ is called the numerator degree of freedom and $(n-k-1)$ is called the denominator degree of freedom. $SSR_r$ is the sum of squared residuals for the restricted model under the null hypothesis and $SSR_{ur}$ is the sum of squared residuals for the unrestricted model under the alternative hypothesis.
					
				\item The level of significance can be chosen as 1\%, 5\% and 10\% and there are corresponding critical values of the $F$ distribution.

                \item Decision rule: If $F_{stats} > F_{critical value}$, the null hypothesis can be rejected at a predetermined level of significance. There is sufficient evidence to conclude that at least one of the explanatory variables’ coefficient does not equal to zero. That is, at least one of the explanatory variable has predictability for future stock returns. Otherwise, the null hypothesis cannot be rejected. This indicates that the overall explanatory variables do not have predictability for future stock returns.
				
			\end{itemize}
			
		\subsubsection{The t-test}
			The t-test can be conducted to find out which explanatory variable should be dropped out from the estimation. The general procedure of implementing the t-test to investigate whether the explanatory variable is statistically significant is as follows according to Wooldridge (2020) \cite{wooldridge2015introductory}:
			
			\begin{itemize}
				\item The t-test has the null hypothesis of the coefficient of an explanatory variable should be equal to zero against an alternative hypothesis of the coefficient should not equal to zero:
					\begin{center}
						$H_0: \theta_j = 0$						
						
						$H_1: \theta_j \neq 0$					
					\end{center}
					and
					\begin{center}
						$H_0: \beta_k = 0$						
						
						$H_1: \beta_k \neq 0$
					\end{center}
					
				\item The test statistic can be calculated as follows:
					\begin{equation}
					\label{tthetahatstatistic}
						t(\hat \theta_j) = \frac{\hat \theta_j}{se(\hat \theta_j)} \sim t_{n-k-1} \ under \ H_0
					\end{equation}
					
					\begin{equation}
					\label{tbetatstatistics}
						t(\hat \beta_k) = \frac{\hat \beta_k}{se(\hat \beta_k)} \sim t_{n-k-1} \ under \ H_0
					\end{equation}
					
				\item The degree of freedom is $(n - k - 1)$, where $n$ is the sample size, $k$ is the number of parameters. The level of significance can be chosen as 1\%, 5\% and 10\% and there are corresponding critical values of the $t$ distribution.

                			\item Decision rule: If $|t(\hat \theta_j)| > |t_{critical value}|$, the null hypothesis can be rejected at a predetermined level of significance. There is sufficient evidence to conclude that the lagged stock return is statistically significant in the multivariate model for stock returns prediction. Otherwise, the null hypothesis will not be rejected and the lagged stock returns are regarded as no predictive power for forecasting stock returns. In this case, $j$ denotes the lag of the stock returns in equation \ref{tthetahatstatistic}.
			
				\item Decision rule: If $|t(\hat \beta_k)| > |t_{critical value}|$, the null hypothesis can be rejected at a predetermined level of significance. There is sufficient evidence to conclude that dividend yield, or earnings-to-price ratio, or book-to-market ratio, or consumption- wealth ratio is statistically significant in the multivariate model for stock returns prediction. Otherwise, the null hypothesis will not be rejected and the explanatory variable is regarded as no predictive power for forecasting stock returns. Different $k$ in equation \ref{tbetatstatistics} denotes dividend yield, earnings-to-price, book-to-market ratio, and consumption-wealth ratio.

			\end{itemize}
			
		\subsection{Out-of-sample Evaluation Method}
		    In this section, the implementation of out-of-sample forecasting will be discussed. Moreover, the Root Mean Squared Error (RMSE) will be introduced as a gauge to measure the out-of-sample forecasting performance.
		
		\subsubsection{Estimation Window}
			There are two commonly used methods for computing estimation window in the out- of-sample forecasting implementation. The first one is the rolling window method and the second one is the recursive (or expansive) window method.
			
			Implementing the rolling window estimation method involves several steps. First, an initial sample period $t = [1, ..., T_1]$ needs to be chosen. Then the forecast period becomes $t = [T + 1, ..., T_1 + T_2]$. Second, the data from $t= [1, ..., T_1]$ are used to estimate the model. Then the one-step ahead out-of-sample forecast $y_{T_1 +1}$ is produced using the estimated parameters in the second step. Any multi-step ahead forecasting can be conducted here. As $Y_{T_1 +1}$ becomes available, the estimation window is moved one period ahead. The third step is to re-estimate the model using $y_S$ from $t = [2, ..., T_1 + 1]$. Then, the one-step ahead out-of-sample forecasts, $y_{T_1 + 1}$ is produced using the estimated parameters in the third step. Any multi-step ahead forecasting can be conducted here. The procedure repeats until the end of the forecast period.
			
			Same as the rolling window method, the recursive scheme also requires to re-estimate the model for each forecast. The steps involved in the recursive scheme are as follows. First, an initial sample period $t = [1, ..., T_1]$ needs to be chosen. Then the forecast period becomes $t = [T_1 + 1, ..., T_1 + T_2]$. Second, the data from $t = [1, ..., T_1]$ are used to estimate the model. Then the one-step ahead out-of-sample forecasts $y_{T_1 + 1}$ is produced using the estimated parameters in the second step. Any multi-step ahead forecasting can be conducted here. The third step is to re-estimate the model using $y_S$ from $t = [1, ..., T_1 + 1]$. This is the main difference between the rolling window and the recursive window. The recursive window will use all the available data to estimate the parameters (Clark, T. E., \& McCracken, M. W., 2009) \cite{clark2009improving}. Then, the one-step ahead out-of-sample forecasts $y_{T_1 + 2}$ is produced using the estimated parameters in the third step. Any multi-step ahead forecasting can be conducted here. The procedure repeats until the end of the forecast period.
			
			Theoretically, the rolling window is easy to implement. However, there are issues in implementing the rolling window estimation method in practice. According to Pesaran and Timmermann (2006) \cite{pesaran1995predictability}, in a situation where a regression model encounters one or more breaks, it can be optimal to use pre-break data to estimate the parameters and then use these parameters to conduct the out-of-sample forecasting. However, in practice, when the time and size of the breaks are unknown, the optimal choice of the estimation window will involve more uncertainties. If the structural breaks affect a particular time series, simply using the full set of pre-break historical data as the estimation window to conduct out-of-sample forecasting could result in large forecast errors. It is difficult to take into account the time and size of the structural change to have an optimal selection of the estimation window if the rolling method is being used.

			Both rolling window and recursive window re-estimate model parameters when there is a new realisation. While the rolling window drops the first observation and shifts the starting point of the estimation window to the next period, the recursive window takes into account both the new realization and the first observation. That is, the recursive scheme expands its estimation window by adding the new realization and use this new set of samples to estimate the parameters for the next forecast. Since the recursive window utilises all the available data to estimate the parameters, it could increase the precision of the in-sample estimation and thus help with providing us with better out-of-sample forecasting performance. Hence, the recursive estimation window method is chosen in this research to conduct the out-of-sample stock returns prediction.
			
		\subsubsection{Root Mean Squared Error (RMSE)}
			To conclude whether a model performs better than the other models or whether a model performs reasonably well, the models’ performance needs to be quantified. Many measures are able to compute a value and provide it as a gauge for evaluating models’ performance. Root Mean Squared Error (RMSE) is one of the measures commonly used to evaluate forecasting performance. According to Hyndman and Koehler (2006) \cite{hyndman2006another}, RMSE is popular because of its theoretical relevance in statistical modeling and it has the same scale as the data. It calculates the differences between the values predicted by the model and the true values. RMSE presents the square root of the quadratic mean of the true values and the predicted values. It can sum up the magnitude of the errors from the prediction and provide a single value of RMSE. When the value of RMSE is large, it indicates that there are large deviations of the values predicted by the model and the true values. In this case, the model has poor forecasting performance. When the value of RMSE is low, it suggests that the values predicted by the model are closed to the true values. This means the model provides good forecasting performance.

			In this research, the RMSE is computed for different models to evaluate their different steps ahead forecasting ability as well as to evaluate the model predictability as a whole. Based on the RMSE results, conclusions can be made on whether the multivariate model could beat the popular historical mean model and the autoregressive model in the out-of-sample stock returns prediction. It is expected that the multivariate model could provide a lower RMSE than the other benchmark models in some periods since the construction of the multivariate model is theoretically reasonable.
			
			The RMSE used in this research is of the form equation \ref{rmse}:
			
			\begin{equation}
			\label{rmse}
				RMSE = \sqrt{\frac{1}{T - [Tr]} \sum \limits _{S = [Tr] + 1}^T (y_S - \hat y_S)^2}
			\end{equation}
			where:
			\begin{itemize}
			    \item $T$ is the total sample periods.
			    
			    \item $[Tr]$ is the in-sample periods.
			    
			    \item $y_S$ is the observed stock returns values.
			    
			    \item $\hat y_S$ is the values predicted by the model (The models include: Multivariate model, Historical mean model, and Autoregressive model which will be discussed in section 4
			    
			\end{itemize}
			
	\section{Data}
	In this section, the source and description of the data used in this study will be presented.
	
	    \subsection{Data Source}
    		In this study, both quarterly and monthly data of S\&P500 price, dividends, earnings, consumption-wealth ratio are used. Price, dividends, earnings, and book-to-market ratio are from the Center for Research in Security Prices (CRSP) dataset. The predicting variables are dividend yield, book-to-market ratio, earnings-to-price ratio as well as consumption-wealth ratio.
    		
    		According to Goyal and Welch (2003) \cite{goyal2003predicting}, dividend yield can be calculated by:
    		\begin{equation}
    		\label{dividendyield}
    			\frac{D_t}{P_{t-1}}
    		\end{equation}
    		where $D_t$ is the total dividends paid by all stocks, and $P_{t-1}$ is the stock price at the beginning of the year.
    		
    		The earnings-to-price ratio is calculated by:
    		\begin{equation}
    		\label{epratio}
    			\frac{E_t}{P_t}
    		\end{equation}
    		where $E_t$ is the total earnings, and $P_t$ is the stock price at the end of the year.
    		
    		The stock return is calculated by:
    		\begin{equation}
    		\label{stockreturn}
    			\frac{P_t - P_{t-1} + D_t}{P_{t-1}}
    		\end{equation}
    		which uses the stock price at the end of the year minus the stock price at the beginning of the year plus total dividends paid divided by the stock price at the beginning of the year. This is a commonly used formula in finance. It considers the source of total stock return is the dividends and its increase in value.
    		
    	\subsection{Data Description}
    	    \subsubsection{S\&P 500 Quarterly Data}
    	        The first dataset being used is the S\&P500 quarterly data. The data are from Quarter 1, 1952 to Quarter 4, 2019 with a total number of 271 available data points.$DY$ denotes dividend yield. $EP$ denotes the earnings-to-price ratio. $Bm$ denotes the book-to-market ratio. $cay$ denotes the consumption-wealth ratio. $SR$ denotes the S\&P500 stock returns. The sample means, standard deviation, skewness kurtosis, and the first-order autocorrelation coefficient are summarised in Table 1.
    	        
    	        From Table 1, it can be seen that there are weak serial correlations in stock returns with the first-order autocorrelation coefficient, which equals 0.094. Dividend yield, earnings-to- price ratio, and book-to-market ratio are highly persistent with the first-order autocorrelation coefficient closed to 1. This is consistent with the existing literature that the financial ratios are extremely persistent. Consumption-wealth ratio is treated as stationary in this study since the variables used to construct the consumption-wealth ratio are presumed stationary according to Lettau and Ludvigson (2001) \cite{lettau2001consumption}. The sample kurtosis is 5.8946 for the stock returns, indicating the fat tail characteristic of financial returns. Fat tail indicates that there is a greater likelihood for extreme events to happen. In the finance content, the fat tail suggests that there is a higher probability that the investment in the stock market could provide extremely high profits or extremely low profits.
    	        
    	        To check for stationarity, the time-series plots of dividend yield, earnings-to-price ratio, book-to-market ratio, and stock returns against time from Quarter 1, 1952 to Quarter 4, 2019 are visually inspected. These plots can be seen in Figure 1. For a variable to be stationary, it needs to satisfy the following three conditions:
    	        \begin{itemize}
    	            \item $1^{st}$ condition: $E(x_t) = \mu$ (a finite constant) for all $t$. This suggests that the mean is time invariant.
    	            
    	            \item $2^{nd}$ condition: $Var(y_t) = E[(x_t - \mu)^2] = \gamma_0$ (a finite constant) for all $t$. This suggests that the variance is also time invariant.
    	            
    	            \item $3^{rd}$ condition: $Cov(x_t, x_{t+j}) = E[(x_t -\mu)(x_{t+j} -\mu)] = \gamma_j$ for all $t$ and $j$. This suggests that the covariance between any two observations depends only on the time interval separating them and not on time itself.
    	        \end{itemize}
    	        
    	        Based on the stationary conditions, only the time-series plot of stock returns against time seems to have a time-invariant mean around 0.05 and a time-invariant variance. It is difficult to locate a mean for the dividend yield, earnings-to-price ratio, and book-to-market ratio in their time-series plots. The covariance for these variables cannot be determined simply based on the graphs. It is hard to conclude whether a variable is stationary or not by visually inspecting the time-series plots. Hence, a statistical test would be utilised which ensures greater reliability in concluding whether the variables are stationarity or not.
    	        
    	        In this research, the ADF test is conducted to confirm whether dividend yield, earnings- to-price ratio, book-to-market ratio are stationary or not. The results of the ADF test are summarised in Table 2. Conclusions are made on whether the times series follow a unit root process based on the p-value. In this research, the significance level is set to be 1\% to increase the precision of the study. If the p-value is greater than 1\%, the null hypothesis can be rejected. There is sufficient evidence to conclude that that the variable does not have a unit root and the variable is stationary. Otherwise, the variable is suggested to be nonstationary. Referring to Table 2, the p-value for dividend yield, earnings-to-price ratio, book-to-market ratio, and stock returns are 0.1283, 0.0344, 0.5367, and 0.0000 respectively. Since all the predictors’ p-value are greater than 1\%, the S\&P500 quarterly data of dividend yield, earnings-to-price ratio, and book-to-market follow a unit rot process, and thus, they are nonstationary. With respect to the stock returns, it computes ADF p-value of 0.0000, the null hypothesis can be rejected. There is sufficient evidence to conclude that that stock returns do not have a unit root at the 1\% level of significance. Hence, stock returns are considered to be stationary.
            
            \subsubsection{S\&P500 Monthly Data}
                In this research, S\&P500 monthly data is also used. The data are from Month 12, 1920 to Month 12, 2019 with a total number of 1185 available data points. $DY$ denotes the dividend yield. $EP$ denotes the earnings-to-price ratio. $Bm$ denotes the book-to-market ratio. $SR$ denotes the S\&P500 stock returns. The sample means, standard deviation, skewness kurtosis, and the first-order autocorrelation coefficient are summarised in Table 3.
                
                From Table 3, it can be seen that there are weak serial correlations in stock returns with the first-order autocorrelation coefficient, which equals 0.157. Dividend yield, earnings-to- price ratio, and book-to-market ratio have the first-order autocorrelation coefficients equal to 0.981, 0.987, and 0.985 respectively. The values of these first-order autocorrelation coefficients are closed to 1, which indicates high persistence. The sample kurtosis of stock returns is 14.8548 in monthly data. The magnitude of this sample kurtosis is considered to be very large compared to the normal distribution, which has a kurtosis of 3. This further confirms the fat tail characteristic of the stock returns.
                
                Similarly, to check for stationarity, the time-series plots of dividend yield, earnings-to- price ratio, book-to-market ratio, and stock returns against the period from Month 12, 1920 to Month 12, 2019 are visually inspected. These plots can be seen in Figure 2. Referring to Figure 2, it seems that it is possible to locate a mean for all variables. However, it is evident that the variance changes across time. Volatility is large especially around the year 1930. This could be due to the Wall Street Crash of 1929 where the share prices on the U.S. stock market collapsed.
                
                However, making a conclusion on whether the variable is stationary or not simply by visual inspection is not reliable. Hence, the ADF test is also conducted for the monthly dataset and the results are summarised in Table 4. According to Table 4, the p-values for dividend yield, earnings-to-price ratio, book-to-market ratio, and stock returns are 0.0234, 0.0187, 0.0276, and 0.0000 respectively. Compared to the quarterly data, the p-values in the monthly setting are relatively lower. However, the dividend yield, earnings-to-price ratio, and book-to- market ratio have p-values that are greater than 0.01. Hence, the null hypothesis cannot be rejected. Thus, dividend yield, earnings-to-price ratio, and book-to-market ratio follow a unit root process. They are considered to be nonstationary. With respect to the stock returns, it has a p-value of 0.0000. This confirms that stock returns are stationary.
                
            \subsubsection{Reduction in the Level of Nonstationarity with Data Visualization}
                According to Appendix 1, since the financial ratios are considered to be extremely persistent, they are assumed to follow a random walk process. Based on Figure 1 and Figure 2, the random walk process assumption is reasonable. The mathematical proof in Appendix 1 shows that the predictors’ variance diverges when time increases. From section 3.2.1 and section 3.2.2, the financial ratios are tested for unit root, which ensures that the variance is not time-invariant. Based on Appendix 1, theoretically, financial ratios’ persistency feature is weakened by the exponential term and thus, the level of nonstationarity will decrease. This section checks whether the level of nonstationarity is indeed reduced when applying real data.

    		    To check whether the data adjusted for the exponential term $(x_{t-1, k} e^{- \frac{x_{t-1, k}^2}{2}})$ has a lower level of nonstationarity than the raw data $(x_{t-1, k})$ their time series plots are inspected first. Figure 1 and Figure 2 are the time series plots of dividend yield, earnings-to-price ratio, and book-to-market ratio for quarterly and monthly data respectively. Figure 3 and Figure 4 show the time series of the quarterly and monthly data, which includes dividend yield, earnings-to-price ratio, and book-to-market ratio, adjusted for the exponential term. However, it is difficult to figure out the difference by simple visual inspection as the graphs are quite similar. This is because the magnitude of the financial ratios is always less than 1. In this case, $e^{- \frac{x_{t-1, k}^2}{2}}$ is also a very small value. Hence, there is only a small difference between $x_{t-1,k}$ and $(x_{t-1, k} e^{- \frac{x_{t-1, k}^2}{2}})$ and the graphs might not be able to present the differences. It is worth noting that, theoretically, $e^{- \frac{x_{t-1, k}^2}{2}}$ will go to zero when $\frac{x_{t-1, k}^2}{2}$ increase. However, this only happens when there are infinitely many $x_{t-1, k}$. In this research, there is a finite sample size and therefore, $e^{- \frac{x_{t-1, k}^2}{2}}$ will not become zero in this study.
    		    
    		    Since Figures 1,2,3 and 4 have small differences, the standard deviation of the first derivative of the time series is used as a proxy for measuring the level of nonstationarity. The first derivative of the time series is the slope, which is an instantaneous rate of change. The standard deviation of the first derivative provides the volatility of the instantaneous rate of change. When the standard deviation is high, it indicates there is always a large variation from one point to another point throughout the entire period. In this case, it is regarded to have a high level of nonstationarity. When the standard deviation is low, it suggests there is a low variation from one point to another point and the time series is considered to be stable. In this case, it is considered to have a low level of nonstationarity. The standard deviation of the first derivative of the time series is plotted in Figure 5 and Figure 6 for quarterly and monthly data respectively. As shown in Figure 5 and Figure 6, dividend yield, earnings-to-price ratio, and book-to-market ratio’s standard deviations of the first derivative first jump to a high value and then decrease as time increases. There are only a few values involved at the beginning and thus the standard deviation of the instantaneous rate of change could be low. When more values are involved, the standard deviation will increase at the beginning. This could be the reason for the jump. As time increases, the standard deviation of the instantaneous rate of change decreases for each financial ratio. This suggests that the volatility of the time series decreases as time increases, which is consistent with the Appendix 1 mathematical proof. Hence, the exponential term did help with reducing the level of nonstationarity in practice.
    		    
    		    In the following section, empirical analysis of the multivariate model will be conducted. The in-sample and out-of-sample sample estimation will be implemented to assess the model’s performance for stock returns prediction.
    		    
    \section{Empirical Analysis}
        \subsection{In-sample Estimation}
            In this section, the in-sample estimation will be implemented using both quarterly and monthly data. The coefficients will be estimated using the Ordinary Least Squared estimation method. According to Inoue and Kilian (2004), when the number of predictors is moderately large relative to the sample size, a testing procedure can be implemented to decide which predictors should be included in the out-of-sample forecasting model and which should be dropped. The testing procedure is as follows. First, fit a model which includes all potentially relevant predictors. Then, the F-test will be conducted to investigate the overall predictability and the t-test will be conducted to check each predictor’s predictability. When the t-test shows that the predictor is insignificant, the predictor can be dropped out before implementing the out-of-sample forecasting. However, such a test could lead to inherently unstable decision rules and may cause inconsistent decisions on whether a predictor should be added or should be dropped under different circumstances. This instability will increase the variance of the stock return predictions and could destroy the accuracy of out-of-sample prediction in practice. The explanatory variables, which include lagged stock returns, dividend yield, earnings-to-price ratio, book-to-market ratio, and consumption-wealth ratio, will be used to conduct the in- sample estimation. The results of the test will be discussed in Section 5.1.2.
        
            \subsubsection{Quarterly and Monthly In-sample Estimation Models}
                The OLS estimation for equation \ref{monthlydata} and equation \ref{monthlynocay} is conducted in this section. S\&P500 quarterly data is applied in equation \ref{monthlydata} while the monthly data is applied in equation \ref{monthlynocay}.
                
                The equation \ref{monthlydata} is the following:
                \begin{equation}
                \label{monthlydata}
                    \hat y = \sum \limits _{j=1}^p \hat \theta_j y_{t-j} + \hat \beta_{cay} x_{t-1, cay} + \sum \limits _{k=1}^q (\hat \alpha_k + \hat \beta_k x_{t-1, k})e^{- \frac{x_{t-1, k}^2}{2}} 
                \end{equation}
                where $x_{t-1, k} = (x_{t-1, DY}, x_{t-1, EP}, x_{t-1, Bm})^{'}$. $DY$ denotes dividend yield. $EP$ denotes earnings-to-price ratio. $Bm$ denotes book-to-market ratio. $cay$ denotes consumption-wealth ratio.
                
                The equation \ref{monthlynocay} is the following:
                \begin{equation}
                \label{monthlynocay}
                    \hat y_t = \sum \limits _{j=1}^p \hat \theta_j y_{t-j} + \sum \limits _{k=1}^q (\hat \alpha_k + \hat \beta_k x_{t-1, k}) e^{- \frac{x_{t-1, k}^2}{2}} 
                \end{equation}
                where $x_{t-1, k} = (x_{t-1, DY}, x_{t-1, EP}, x_{t-1, Bm})^{'}$. $DY$ denotes dividend yield. $EP$ denotes earnings- to-price ratio. $Bm$ denotes book-to-market ratio. Since the monthly data for consumption- wealth ratio is not available, consumption-wealth ratio is excluded from the monthly in-sample estimation.
                
            \subsubsection{Quarterly and Monthly In-sample Estimation Outputs}
                The quarterly in-sample estimation outputs are summarised in Table 5. The monthly in-sample estimation outputs are summarised in Table 6. The adjusted R-squared for quarterly estimation is 0.05601 and for monthly estimation is 0.1236. These two adjusted R-squared values are quite low, indicating that the variables included might not help with adding value to the model. However, both quarterly and monthly estimations have F-statistics with an extremely low p-value. The p-value for quarterly estimation is 0.006825 and the p-value for monthly estimation is 2.2e-16. Both p-values are less than 1\%. In this case, the null hypothesis of all the predictors’ coefficients should be equal to zero can be rejected at 1\% level of significance. There is sufficient evidence to conclude that at least one of the predicting variables in the model is significant.
                
                According to Lamont (1998) \cite{lamont1998earnings}, high dividends forecasts high returns, and high earnings forecast low returns. In this study, using S\&P500 data, the estimated coefficient for dividend yield is positive in both quarterly and monthly estimations, which is consistent with Lamont’s study. The estimated coefficient for the earnings-to-price ratio is positive in quarterly estimation but negative in monthly estimation.
                
                Three-period lagged stock return has a p-value of 0.0275 in quarterly estimation. If a judgment is made solely based on the p-value for the variables’ predictability, $y_{t-1}$ is the only variable that has predictability. With respect to monthly estimation, one-period lagged dividend yield, which has a p-value of 0.000589, and two-period lagged stock return, which has a p- value of 0.006079, have predictability. The dividend yield in monthly estimation has the same result as Goyal and Welch (2003) \cite{goyal2003predicting} where the dividend yield retains good statistical significance in the in-sample estimation.
                
                However, conventional tests of the predictability of stock returns could be invalid when the predictor is persistent Campbell et al., 2006 \cite{campbell2006efficient}. Financial ratios are persistent but they are also proven to have predictability. Thus, conclusions should not be made for whether the variables have predictability or not solely based on the p-value. In addition to that, in-sample predictability does not necessarily equal out-of-sample predictability. Using financial ratios to predict stock returns is controversial where the in-sample and out-of-sample estimations are inconsistent according to Lettau et al., 2008 \cite{lettau2008reconciling}. The variables could act differently in the in-sample and out- of-sample estimation. In this research, the subject of interest is stock returns prediction, and thus, the main focus is on out-of-sample evaluation, which will be discussed in the following section 4.2.
                
        \subsection{Out-of-Sample Evaluation}
            In this section, the out-of-sample estimation will be conducted to evaluate the forecasting performance of the multivariate model. Besides the multivariate model, the historical mean model and autoregressive model will be also included as a comparison.
            
            The reasons for choosing the historical mean model as a benchmark are as follows. In the existing studies, most evidence supports the in-sample predictability using those popular financial ratios. With respect to out-of-sample prediction, Goyal and Welch (2008) \cite{welch2008comprehensive} argue that the out-of-sample stock return prediction is unable to overcome the historical mean model forecasting. Campbell and Shiller (2001) \cite{campbell2001valuation} find that this is particularly during the bull market of the late 1990s. However, Goyal and Welch (2008) \cite{welch2008comprehensive} consider that it is a systematic problem where the in-sample correlations conceal a failure of the out-of-sample prediction and are not confined to any specific periods. Several predicting regressions are compared with the historical mean model to predict stock returns. It is found that the historical mean model always computes remarkable stock return forecasts. Therefore, Goyal and Welch (2008) \cite{welch2008comprehensive} conclude that “the profession has yet to find some variable that has meaningful and robust empirical equity premium forecasting power”. This research will face the challenge from Goyal and Welch (2008) \cite{welch2008comprehensive} to check if the new multivariate model could have a better out-of-sample performance than the historical mean model using those popular financial ratios as predictors.
            
            With respect to the autoregressive model, it is a model that only consists of lagged values of stock returns. In this research, the lagged period is up to four. When the quarterly data are used, there are one-quarter ahead forecasts, two-quarter ahead forecasts, three-quarter ahead forecasts, and four-quarter ahead forecasts. When the monthly data are used, there are one-month ahead forecasts, two-month ahead forecasts, three-month ahead forecasts, and four- month ahead forecasts. This autoregressive model is a benchmark model in this research. This is because if the autoregressive model can compute better out-of-sample performance than the multivariate model, it suggests that the financial ratios as well as the consumption-wealth ratio have little out-of-sample predictability. This will be consistent with the results shown in Goyal and Welch (2008) \cite{welch2008comprehensive}. Moreover, the exponential term which is used to reduce the level of nonstationarity from the predictors is meaningless. In this case, the new multivariate model does not contribute to the stock returns prediction.
            
            To compare different models’ out-of-sample forecasting performance, RMSE is used. The contribution from this research is to show that the new multivariate model including financial ratios and consumption-wealth ratio is possible to perform better than the historical mean model and autoregressive model in the out-of-sample prediction. In this study, S\&P500 quarterly and monthly data are used to conduct the out-of-sample forecasting. Multi-step ahead forecasting is also implemented for each model, which will be evaluated according to RMSE.
            
            \subsubsection{Model Specification for Out-of-sample Evaluation}
                For S\&P500 quarterly data, the predictive regression relationship is estimated using the following four models:
                \begin{center}
                (Model 1-1) Multivariate model with consumption-wealth ratio, which is equation \ref{multivariatemodelformadjusted}.
                \end{center}
                
                (Model 1-2) Historical mean model:
                \begin{center}
                    $y_t = \mu + e_t$
                \end{center}
                Since Goyal and Welch (2008) \cite{welch2008comprehensive} argue that the out-of-sample stock return prediction is unable to overcome the historical mean model forecasting, this research would want to face this challenge and to check if the same situation happens. Thus, the historical mean model is included in this research as a benchmark model.
                
                (Model 1-3) Autoregressive AR(4) model:
                \begin{center}
                    $y_t = \theta_1 y_{t-1} + \theta_2 y_{t-2} + \theta_3 y_{t-3} + \theta_4 y_{t-4} + e_t$
                \end{center}
                The AR(4) model is also introduced as one of the benchmark models. The difference between the AR(4) model and the multivariate model (Model 1-1) is whether the predictors such as dividend yield, earnings-to-price ratio, and book-to-market ratio are included in the model or not. If this AR(4) model has better out-of-sample forecasting performance than Model 1-1, it can be concluded that the predictors including consumption-wealth ratio, dividend yield, earnings-to-price ratio, and book-to-market ratio do not help with adding predictability to Model 1-1.
                
                (Model 1-4) Multivariate model without consumption-wealth ratio:
                \begin{center}
                    $y_t = \sum \limits _{j=1}^p \theta_j y_{t-j} + \sum \limits _{k=1}^q (\alpha_k + \beta_k x_{t-1, k}) e^{- \frac{x_{t-1, k}^2}{2}} + e_t$
                \end{center}
                Model 1-4 is the third benchmark model in out-of-sample forecasting using S\&P500 quarterly data. The variables included in Model 1-4 have similar settings as Model 1-1. The only difference is Model 1-4 does not include that consumption-wealth ratio as a predictor. According to Lettau and Ludvigson (2001) \cite{lettau2001consumption}, the consumption-wealth ratio is a strong predictor for future stock returns prediction. If Model 1-4 can outperform Model 1-1 in the out-of-sample forecasting, it means that the consumption-wealth ratio might not have the predictability, which is not aligned with Lettau and Ludvigson’s findings \cite{lettau2001consumption}.
                
                For S\&P500 monthly data, the predictive regression relationship is estimated using the following three models:
                \begin{center}
                (Model 2-1) Multivariate model with consumption-wealth ratio:
                    $y_t = \sum \limits _{j=1}^p \theta_j y_{t-j} + \sum \limits _{k=1}^q (\alpha_k + \beta_k x_{t-1, k}) e^{- \frac{x_{t-1, k}^2}{2}} + e_t$
                \end{center}
                The variables in Model 2-1 have the same settings as Model 1-1 except that Model 2-1 does not include the consumption-wealth ratio as a predictor. This is because the S\&P500 dataset does not have available monthly consumption-wealth ratio series. In this case, $x_{t-1, k}$ includes dividend yield $(k = 1)$, earnings-to-price ratio $(k = 2)$, and book-to-market ratio $(k = 3)$. With regard to $y_{t-j}, j = 1,2,3,4$. However, the one period in Model 2-1 indicates one month which is different from the settings in Model 1-1. $j$ does not take the values up to twelve to involve the past one-year information into the model. This is because having $j$ from one to twelve, which suggests twelve predictors in the model, could lead to an overfitting problem. Thus, having $j$ from one to four, which involves the past one-quarter information into the model, is relatively more reasonable in conducting the monthly out-of-sample evaluation.
                
                (Model 2-2) Historical mean model:
                \begin{center}
                    $y_t = \mu + e_t$
                \end{center}
                Model 2-2 is the first benchmark model in using the S\&P500 monthly data for out-of- sample stock returns prediction. The reason for choosing this model is the same as Model 1-2.
                
                (Model 2-3) Autoregressive AR(4) model:
                \begin{center}
                    $y_t = \theta_1 y_{t-1} + \theta_2 y_{t-2} + \theta_3 y_{t-3} + \theta_4 y_{t-4} + e_t$
                \end{center}
                Model 2-3 is the second benchmark model in using the S\&P500 monthly data for out- of-sample stock returns prediction. The reason for choosing Model 2-3 as a benchmark model is the same as Model 1-3.
                
                The following section is to figure out whether the multivariate model has predictability for future stock returns, and thus, the multivariate model’s forecasting performance is compared with the benchmark models. RMSE is used as a gauge for measuring out-of-sample forecasting performance. More information about the RMSE will be discussed in the following.
                
            \subsubsection{RMSE in Out-of-sample Evaluation}
                Based on equation \ref{rmse} in Section 2.5.2, RMSE is computed in two perspectives for out-of-sample forecasting performance. The first perspective is when the multi-step ahead forecasting is implemented, RMSE is calculated for each step and each model. In this case, models’ multi-step ahead predictability can be evaluated. The other perspective is computing the RMSE for each model by taking into account the forecasting from all number of steps ahead. In this case, models’ overall forecasting performance can be evaluated. The results of RMSE will be discussed in Section 4.2.4 for quarterly data and Section 4.2.5 for monthly data.
                
            \subsubsection{Implementation of Recursive Window}
                According to section 2.5.1, the details of implementing the recursive window in this research are provided as the following.
                
                1. For the case of $r=1$:
                
                The first step is to construct the first window where $r = 1$. At the first window, it is standing at the point $x_n$ with $(n-1)$ pairs of observations. The $(n - 1)$ pairs of observations include $\{(x_1, y_2), (x_2, y_3), ..., (x_{n-1}, y_n)\}$. The reason for the different subscript in each pair of observation for $x$ and $y$ is because the multivariate model equation \ref{multivariatemodelform} uses lagged values for stock return prediction.
                
                Use the $(n - 1)$ pairs of observations to predict $y_{n+1}$ and then $\hat y_{n+1}$ can be obtained as the predicted value. $\hat y_{n+1}$ is also called the one-step ahead forecast where $j=1$.
                
                Then, include this new realization $\hat y_{n+1}$ and now there is the new $(n-1)$ pairs of observations which include $\{(x_1, y_3), (x_2, y_4), ..., (x_{n-1}, \hat y_{n+1})\}$. Use the new $(n-1)$ pairs of observations to predict $y_{n+1}$ and $\hat y_{n+2}$ can be obtained as the predicted value. $\hat y_{n+2}$ is also called the two-step ahead forecast where $j=2$.
                
                Similarly, use the updated $(n-1)$ pairs of observations to predict $y_{n+3}$ and $\hat y_{n+3}$ can be obtained as the predicted value. $\hat y_{n+3}$ is also called the three-step ahead forecast where $j=3$.
                
                It is worth noting that the multi-step ahead forecasting mentioned above are all conducted in the first window $(r=1)$, standing at the point $x_n$. When the S\&P500 quarterly data is used, there is a total number of 271 available data points, and the out-of-sample forecasting starts at time index $n=200$. That is, the initial in- sample period for estimating the parameters is 200. At the first window $(r=1)$, it is standing at the point $x_200$ to predict $y_201, y_202, y_203, y_204$ for $j = 1, 2, 3, 4$ respectively. Then the multi- step ahead forecasting using S\&P500 quarterly data in the first window is considered to be complete.

                For the S\&P500 monthly data, there is a total number of 1185 available data points, and the out-of-sample forecasting starts at time index $n=948$. That is, the initial in-sample period for estimating the parameters is 948. The out-of-sample period starting point is calculated by 80\% $\times$ 1185 = 948. 80\% of the sample is used to estimate the parameters before conducting out-of-sample forecasting. At the first window $(r = 1)$, it is standing at the point $x_948$ to predict $y_949, y_950, y_951, ..., y_960$ for $j=1, 2, 3, ..., 12$ respectively. Then the multi-step ahead forecasting using S\&P500 monthly data in the first window is considered to be complete.
                
                2. For the case of $r =2$:
                
                After the multi-step ahead forecasting is completed in the first window, the second window is constructed $(r=2)$. At the second window, it is standing at the point $x_{n+1}$ with $n$ pairs of observations. The $n$ pairs of observations include $\{(x_1, y_2), (x_2, y_3), ..., (x_n, y_{n+1})\}$. The sample size for estimation has been expanding until this last window and the observations include until $(x_{n+R-2}, y_{n+R-1})$ now. These $(n+R)$ pairs of observations are then used to predict $y_{n+R}, y_{n+R+1}, ..., y_{n+R+j-1}$.
                
                When the S\&P500 quarterly data is used, the last window $R$ is equal to 68. This is calculated by $R = 271 - 200 - 4 + 1 = 68$ where the entire sample size $N = 271$, initial in- sample size $n = 200$, and there are up to four-step ahead forecasting $(j = 1, 2, 3, 4)$. $R = 68$ to ensure that the quarterly data out-of-sample time periods are fully predicted. When $R = 68$, it is standing at the point $X_267$ with 268 pairs of observations to predict $y_{268}, y_{269}, y_{270}, y_{271}$ for $j = 1, 2, 3, 4$ respectively. A full dataset of the out-of-sample predicted values can be collected from $r = 1, 2, 3, ..., 68$ for $j = 1, 2, 3, 4$ respecitvely.
                
                When the S\&P500 monthly data is used, the last window $R$ is equal to 226. It can be calculated by $R = 1185 - 948 - 12 + 1 =226$ where the entire sample size $N = 1185$, initial in-sample size $n = 948$, and there are up to twelve-step ahead forecasting $(j = 1, 2, ..., 12)$. $R = 226$ ensures that the S\&P500 data out-of-sample time periods are fully predicted. When $R = 226$, it is standing at the point $x_1173$ which include 1172 pairs of observations to predict $y_{1174}, y_{1175}, ..., y_{1285}$ for $j = 1, 2, ..., 12$ respectively. A full dataset of the out-of-sample predicted values can be collected from $r = 1, 2, 3, ..., 226$ for $j = 1, 2, ..., 12$ respecitvely.

                Overall, the recursive window is used for the multivariate model, historical mean model, and the autoregressive model for S\&P500 quarterly and monthly data. Based on the above discussion on the implementation of the recursive window estimation method, the out-of-sample prediction dataset can be collected and the general information is as follows. For quarterly data, there are $j$ = 1,2,3,4 ahead forecasting and for each $j$, there are 66 predicted values. Thus, there is a total number of $68 \times 4 = 272$ stock returns predicted values for each model.
                
                For monthly data, there are $j$ = 1,2,3, ... ,12 ahead forecasting and for each $j$, there are 226 predicted values. Thus, there is a total number of $226 \times 12 = 2712$ stock returns predicted values for each model.
                
                The results of the multi-step ahead stock returns forecasting and the values of RMSE calculated based on the forecasting results will be shown in section 4.2.4 for quarterly data and section 4.2.5 for monthly data.

            \subsubsection{Quarterly Data Out-of-sample Predicting Outputs Evaluation}
            
                1. For the case of $j=1$:
                
                The situation when $j$ = 1 will be discussed firstly, the one-step ahead forecasting. From Figure 7, it can be seen that there are a few large deviations from the mean of the true stock returns. The first big deviation is at around the time index 0 to 8, corresponding to the period from 2000 to 2003. The deep drops of the S\&P500 stock returns from 2000 to 2003 could be due to the September 11 attacks in 2001, and the energy crisis that happened in the 2000s. The U.S. stock market was hit hard by these events at that time. Due to the September 11 attacks, the stock exchange had to close on September 11 and remain closed for almost a week. For the 2000s energy crisis, there was a steep rise in oil prices at that period. The high oil prices and weak economy led to demand contraction. In this case, the stock market was also affected.
                
                From Figure 7, Model 1-2, which is the historical mean model, is a horizontal straight line through the mean of the S\&P500 stock returns. When the S\&P500 stock returns are stable, there are some overlapping of the historical mean model and the true stock returns. Model 1-2 does not react to any of the events which were mentioned at the beginning of this section.
                
                Model 1-3, which is the AR(4) model, seems to compute a good forecasting performance since the pattern is similar to the true S\&P500 stock returns. However, if the graph is checked more carefully, it can be seen that when the stock returns start increasing, AR(4) model predicts the return will increase even further in the next period. When the stock returns start decreasing, AR(4) model predicts the return will decrease further in the future. The AR(4) model has over predicted stock returns. That is, the predicted values have a larger magnitude compared to the true stock returns. This is because AR(4) model only includes lagged stock returns as predictors and thus, it will follow the true S\&P500 stock returns track for its future prediction. AR(4) model is sensitive to any economic crises, however, the volatility for AR(4) is too large. For example, AR(4) model predicts a much lower stock return than the true one during the global financial crisis period, indicating large forecast errors. Thus, it is expected that the RMSE for the AR(4) model will be relatively high.
                
                Model 1-1 and Model 1-4 are multivariate models. Model 1-1 is the multivariate model with consumption-wealth ratio, which is presented in black line. Model 1-4 is the multivariate model without consumption-wealth ratio without consumption-wealth ratio, which is presented in the green line. Model 1-1 and Model 1-4 have a similar pattern in stock returns prediction. Compared to AR(4) model, Model 1-1 and Model 1-4 are relatively less volatile. From Figure 7, it can be seen that both Model 1-1 and Model 1-4 react to the economic crises but the magnitude of increase or decrease due to the crises is less than AR(4). The difference between Model 1-1 and Model 1-3 is that Model 1-1 includes the predictors with the exponential term, and this exponential term is considered to contribute to providing us with a lower level of volatility. If Model 1-1 and Model 1-4 have lower RMSE, it is considered that the financial ratios and the macroeconomic ratio add predictability to our model.
                
                To check the forecasting performance more intuitively, the RMSE is computed. The RMSE for one-step ahead forecasts is calculated by the following equation \ref{68rmse}:
                \begin{equation}
                \label{68rmse}
                    RMSE_{quarterly, j = 1} = \sqrt{\frac{1}{68} \sum \limits _{1}^{68} (y_S - \hat y_{s, j=1})^2}
                \end{equation}
                
                The results of RMSE for one-step ahead forecasts are shown in Table 7. Since the lower the RMSE indicates better forecasting performance, Model 1-4, which is the multivariate model without consumption-wealth ratio, is considered to perform better than all the other models. Model 1-1, which is the multivariate model with consumption-wealth ratio, has the second-lowest value of RMSE. That is, the multivariate structure can beat the historical mean model and autoregressive model in the one-step ahead forecasting using S\&P500 quarterly data.
                
                Model 1-1 has a higher RMSE value than Model 1-4, which is different from the expectation that having the consumption-wealth ratio as a predictor could add predictability to our multivariate model. However, it should not be concluded that the consumption-wealth ratio does not have predictability for future stock returns solely based on one simple data series for $j = 1$. $j = 2, 3, 4$ will be also discussed to check whether the multivariate model structure performs better than other models and whether the predictors in Model 1-1 have predictability.
                
                2. For the case of $j = 2$:
                
                Figure 8 shows two-step ahead forecasting for all the models and Table 7 shows the RMSE results for two-step ahead forecasts. The RMSE is calculated in the same way as equation (18) but in the case of $j=2$. From Figure 8, Model 1-1, Model 1-3, and Model 1-4 are less volatile and have a similar pattern as Model 1-2, the historical mean model. This is different from the one-step ahead forecasting. It is possible for the two-step ahead forecasting to be less precise than one-step ahead forecasting. For example, it is hard to know the global financial crisis will happen two quarters earlier and hard to predict how bad the situation can be. Even though this is the case, Model 1-1 is still able to compute the lowest value of RMSE, which is 0.07381919 from Table 7.
                
                As mentioned in section 2, the consumption-wealth ratio is considered to have predictability at the short and intermediate horizons. In this case, the two-step ahead forecasting means two-quarter ahead forecasting. Model 1-1 beats Model 1-4 and this is because the two- quarter time frame would allow for the consumption-wealth ratio to take effect. In general, Model 1-1 has better out-of-sample forecasting performance compared to all the other benchmark models in terms of the magnitude of RMSE.
                
                3. For the case of $j = 1, 2, 3, 4$ collectively:
                
                Besides checking RMSE for individual $j = 1, 2, 3, 4$, the RMSE for $j = 1, 2, 3, 4$ forecast results is also calculated collectively and it is computed in Table 8. The RMSE is calculated by the following equation :
                \begin{equation}
                \label{RMSEj1234}
                    RMSE_{Quarterly, j = 1, 2, 3, 4} = \sqrt{\frac{1}{272} \sum \limits _{1}^{272} (y_S - \hat y_{s, j = 1, 2, 3, 4})^2}
                \end{equation}
                Recall that 272 is the total number of predicted values and the details of it were provided in section 4.2.3.
                
                It is expected that Model 1-1 should provide better out-of-sample forecasting performance in general, compared to all other benchmark models. The reasons are as follows. Firstly, lagged stock returns are included as predictors, and thus, Model 1-1 includes all the information that AR(4) model has. Secondly, Model 1-1 includes financial ratios as predictors. Furthermore, the level of nonstationarity due to the financial ratios is reduced by introducing an exponential term. The financial ratios should add predictability to Model 1-1. Lastly, the consumption-wealth ratio, which is considered a strong predictor for stock returns prediction, is also included in Model 1-1.
                
                From Table 8, Model 1-1 has the better overall performance compared to Model 1-2, Model 1-3, and Model 1-4, in terms of the magnitude of RMSE. This is consistent with the expectation mentioned above. Overall, if models are evaluated for their overall forecasting performance, taking into account all the results from the multi-step ahead forecasting, Model 1-1 has better out-of-sample forecasting performance than all the other benchmark models.
                
            \subsubsection{Monthly Data Out-of-sample Predicting Outputs Evaluation}
            
                1. For the case of $j = 1$:
                
                The out-of-sample prediction starts from time index 1041 to 1185 which corresponds to predicting periods from around 2000 to 2019. The reasons for the large deviations in the true stock returns are the same as what were mentioned in section 4.2.4.
                
                The one-step ahead forecasting $(j=1)$, will be discussed first. In the monthly settings, one step means one month. Referring to Figure 11, Model 2-1, which is the historical mean model, has a similar graph as Figure 7. The monthly data has more fluctuations and the historical mean model does not seem to be able to capture the features of the true stock returns. Moreover, Model 2-3, which is the AR(4) model, has more fluctuations than in the quarterly out-of-sample forecasting. Again, Model 2-3 over predicts the true stock returns. It can be also seen from Figure 11 that the AR(4) model’s forecast errors and volatility are very large. With respect to Model 2-1, it has relatively low volatility than AR(4) model. The forecast errors from Model 2-1 seem to be less than the AR(4) model. Whether Model 2-1 has a better out-of-sample one- step ahead forecasting performance can be checked by RMSE. Similar to equation \ref{68rmse}, the RMSE is calculated for one-step ahead forecasting using S\&P500 monthly data as follows :
                \begin{equation}
                \label{rmse226}
                    RMSE_{Monthly, j =1} = \sqrt{\frac{1}{226} \sum \limits _{1}^{226}(y_S - \hat y_{s, j =1})^2}
                \end{equation}
                Recall that for each $j$, there is a total of 226 predicted values for each model and the details of the calculations for 226 are provided in section 4.2.3.
                
                The RMSE results for the one-step ahead forecasting are summarised in Table 9. It can be seen that Model 2-3 has the largest value of RMSE (0.1163453), indicating the least precision for stock returns prediction. This aligns with expectations that the AR(4) model is the most volatile amongst all three models based on Figure 11. Model 2-1 has the lowest value of RMSE (0.03868196) and Model 2-2 has the RMSE of 0.04490874. Model 2-1 has much lower in terms of the magnitude of RMSE than the other two models, suggesting strong predictability for stock returns prediction in one-step ahead forecasting when using S\&P500 monthly data.
                
                2. For the case of $j =2, ..., j = 12$:
                
                For $j = 2$ to $j = 12$, the out-of-sample prediction outputs are shown in Figure 12 to Figure 22 respectively. Since their prediction outputs are very similar, the implications will be discussed collectively.
                
                Different from Figure 11, Figure 12 to Figure 22 show that AR(4) model’s predictions are less volatile for two-step ahead forecasting to twelve-step ahead forecasting. It is expected that the precision in the predictions for stock returns will increase and thus, the RMSE value for $j = 2$ to $j = 12$ should be less than the RMSE of $j = 1$ with respect to Model 2-3. Model 2-2, which is the historical mean model, does not vary much through $j = 2$ to $j = 12$. Referring to Figures 20,21,22, it can be noticed that from $j = 10$ to $j = 12$, Model 2-2 and Model 2-3 are overlapping. It is expected that the RMSE for $j = 10$ to $j = 12$ computed by Model 2-2 and Model 2-3 will be roughly the same. It is regarded that AR(4) model is losing its predictability for future stock returns when higher order of multi-step ahead forecasting is conducted. With respect to Model 2-1, it retains its reaction and sensitivity to economic crises. The increase and decrease in stock returns predicted by Model 2-1 are roughly the same as the true S\&P500 stock returns. Model 2-1 is not as volatile since there is the exponential term that helps with controlling the volatility. Figure 12 to Figure 22 show that Model 2-1 has stable out-of-sample forecasting performance. For various degrees of forecast horizons, Model 2-1’s predictions are similar to each other.
                
                To investigate the out-of-sample forecasting performance more intuitively, RMSE needs to be calculated. The RMSE is calculated using equation (19) but in the case of $j = 2, 3, 4, ..., 12$ respectively and the results are presented in Table 9 to Table 11. It can be seen that the lowest values of RMSE are always computed by Model 2-1. That is, in any case of $j = 2, 3, 4, ..., 12$, Model 2-1 always presents the best out-of-sample forecasting performance compared to Model 2-2 and Model 2-3, using S\&P500 monthly data.
                
                Moreover, AR(4) model’s RMSE increases from around 0.043 to 0.044 for $j = 2$ to $j = 12$. This suggests that the predictability of this model for future stock returns prediction decreases when higher order of multi-step ahead forecasting is conducted. Also, referring to Table 9,10,11, Model 2-2 and Model 2-3 compute almost the same values for RMSE. This indicates that the Model 2-2’ and Model 2-3’s out-of-sample forecasting performance is roughly the same for $j = 10, 11, 12$. 
                
                In general, Model 2-1 is considered to have strong multi-step ahead forecasting ability for stock returns prediction when S\&P500 monthly data is used. It has better out-of-sample forecasting performance than the historical mean model and autoregressive model in the $j = 1, 2, 3, ..., 12$ multi-step ahead forecasting.
                
                3. For the case of $j = 1, 2, 3, ..., 12$ collectively:
                
                Besides checking the RMSE for the individual $j = 1, 2, 3, 4, ..., 12$, the RMSE for $j = 1, 2, 3, 4, ..., 12$ forecast results is also calculated collectively and the results are summarised in Table 12. The RMSE is calculated by the following equation \ref{rmse2712}:
                \begin{equation}
                \label{rmse2712}
                    RMSE_{Monthly, j=1,2,3,4, ..., 12} = \sqrt{\frac{1}{2712} \sum \limits _{1}^{2712} (y_S - \hat y_{s, j = 1, 2, 3, 4, ..., 12})^2}
                \end{equation}
                Recall that 2712 is the total number of predicted values and the details of calculations for 2712 were provided in section 4.2.3.
                
                From Table 12, it can be seen that Model 2-1 has RMSE of 0.04247248, while Model 2-2’s RMSE is 0.04446577 and Model 2-3’s RMSE is 0.05380378. The multivariate model has the lowest value of RMSE. In this case, it can be concluded that if models are evaluated for their overall forecasting performance, taking into account all the results from the multi-step ahead forecasting collectively, Model 2-1 has better out-of-sample forecasting performance than all the other benchmark models. This conclusion is based on the S\&P500 monthly dataset.
                
    \section{Conclusion and Findings}
        In summation, this research has shown the construction of a new multivariate model which includes dividend yield, earnings-to-price ratio, book-to-market ratio as well as consumption-wealth ratio as explanatory variables, for future stock returns predictions. Since these financial ratios are considered to be persistent, an exponential term is introduced to reduce the level of nonstationarity and hence, increase predictive power. This research has assessed that the level of nonstationarity is indeed reduced by the exponential term. This research has also shown through empirical analysis that the new multivariate model can beat other benchmark models such as the historical mean model and autoregressive model in stock returns prediction most of the time.
        \begin{itemize}
            \item With the mathematical proof and the real data application, the exponential term $(e^{- \frac{x_{t-1,k}^2}{2}})$ is effective in reducing the level of nonstationarity due to the financial ratios. The exponential term is then able to balance the equation which has stationary variable, the stock returns on the RHS, and nonstationary variables, the financial ratios on the LHS.
            
            \item The in-sample estimation suggests that the dividend yield has good in-sample predictability, which is in line with the existing studies. But the new multivariate model does not show good in-sample predictability for stock returns.
            
            \item With respect to out-of-sample evaluation, the multivariate model has satisfactory forecasting performance. The dividend yield, earnings-to-price ratio, book-to- market ratio, and consumption-wealth ratio add predictive power for out-of-sample stock returns prediction when these explanatory variables are included in the multivariate model. Since in general, the new multivariate model beats the autoregressive model, which does not include these predictors.
            
            \item For quarterly out-of-sample prediction, the multivariate model has good predictability for one- and two-step ahead forecasting.
            
            \item For monthly out-of-sample prediction, the multivariate model beats other benchmark models for forecasting any number of steps ahead. In this research, the step $j$ is up to twelve.
            
            \item If the multivariate model is evaluated by the overall forecasting performance (considering $j = 1, 2, 3, 4$ collectively in quarterly case, and considering $j = 1, 2, 3, ..., 12$ colletively in monthly case), it always computes the lower value of RMSE compared to the benchmark models. This suggests that the multivariate model has better out-of-sample forecasting performance than other benchmark models.
            
            \item Overall, this new multivariate model is a reasonably good model for future stock return prediction using monthly and quarterly S\&P500 datasets and is capable of short, intermediate, and long-horizon stock return prediction.
        \end{itemize}
        
        The methodology in this research can be applied to other sets of data that have similar characteristics to the returns and the explanatory variables. The results and findings from this research might have some implications in practice for the investors as well as other participants in the stock markets. This research helps to identify the explanatory variables which contribute to stock returns prediction and this new multivariate model can provide the practitioner with an implication of risk management and risk hedging.
        
        The future scope of this research could involve the following aspects. Firstly, besides the explanatory variables used in this research, there are also many other explanatory variables such as dividend-price ratio, interest rate, and inflation rate. They can also be applied in this new multivariate model and tested for their predictability for stock returns. Secondly, future research can try to find out which combination of the predictors provides the strongest predictive power. Lastly, future studies can extend to any transformation of the multivariate model. For example, adding the interaction terms of the financial ratios to take into account other possible information.
        \\
        
        \textbf{Acknowledgement}: I am extremely grateful to my esteemed supervisor Professor Jiti Gao for all the help, invaluable advice, continuous support, and patience during this research project, without which, I would not have been able to complete this study. I would also like to thank Dr. Tingting Cheng and Miss Ying Zhou for providing me with the technical support and dataset.
        \\
        
        \textbf{Conflicts of Interest}: The authors declare no conflict of interest.
        
    \section{Appendix}
        \subsection{A Brief Mathematical Proof for the reduction in level of nonstationarity}
            Since the financial ratios are considered to be persistent, which implies unit root. They are considered to be nonstationary. To reduce the level of nonstationarity due to the persistent variables, exponential term $(e^{- \frac{x_{t-1,k}^2}{2}})$ is introduced. In this section, a brief mathematical proof is provided. The second moment is calculated here as an implication for the level of nonstationarity.
            
            Regard $X_t$ follows a random walk $(X_t = X_{t-1} + \mu_t)$, since the predictor variables, especially for financial ratios, could be extremely persistent. The variance of $X_t$ can be shown as follows:
            \begin{center}
                $Var(X_t) = Var(\mu_1 + \mu_2 + \cdots \ u_t) = \sigma_{\mu}^2 + \sigma_{\mu}^2 + \cdots + \sigma_{\mu}^2 = t \sigma_{\mu}^2$
            \end{center}
            
            The variance diverges as time increases and thus $X_t$ is nonstationary.
            
            Let $\mu = x_{t-1, k}$. Since $x_{t-1, k}$ follow  random walk, there is $x_t = x_0 + \sum _{i=1}^t \mu_i$, and thus $E(x_t) = 0$ and $x_t \sim N(0, t \sigma_{\mu}^2)$ (assuming $assuming \sigma_{\mu}^2 = 1$ for calculation simplicity):
            
            \begin{itemize}
                \item Before the exponential term $(e^{- \frac{x_{t-1,k}^2}{2}})$ is introduced, the second moment related to the predictors is calculated by the following:
                \begin{align*}
                    E(\alpha_k + \beta_k x_{t-1, k})^2 &= E(\alpha_k^2 + \beta_k^2 x_{t-1, k}^2 + 2 \alpha_k \beta_k x_{t-1, k}) \\
                     &= \alpha_k^2 + \beta_k^2 E(x_{t-1, k}^2) \\
                     &= \alpha_k^2 + \beta_k^2 t \sigma_{\mu}^2
                \end{align*}
                When $t \rightarrow \infty$, $\alpha_k^2 + \beta_k^2 t \rightarrow \infty$. That is, the variance of the model will diverge as time increases. This suggests that the level of nonstationarity is high.
                
                \item After the exponential term $(e^{- \frac{x_{t-1,k}^2}{2}})$ is introduced, the second moment related to the predictors is calculated by the following:
                \begin{align*}
                    E(e^{- \frac{x_{t-1,k}^2}{2}}) &= E(e^{-x_{t-1,k}^2})\\
                    &= \int_{-\infty}^{+\infty} e^{-\mu^2} \frac{1}{\sqrt{2 \pi t}} e^{\frac{\mu^2}{2t}} d\mu \\
                    &= \frac{1}{\sqrt{2 \pi t}} \int_{-\infty}^{+\infty} e^{-\frac{(2t+1)\mu^2}{2t}} d\mu \\
                    &= \frac{1}{\sqrt{2 \pi t}} \sqrt{2 \pi \sigma} \int_{-\infty}^{+\infty} \frac{1}{\sqrt{2 \pi \sigma}} e^{-\frac{\mu^2}{2 \sigma^2}}\\
                    &= \frac{1}{\sqrt{2 \pi t}} \times \sqrt{2 \pi} \sqrt{\frac{t}{2t+1}}\\
                    &= \frac{1}{\sqrt{2t+1}}\\
                \end{align*}
                
            \end{itemize}
            
            The second moment of the exponential term goes to zero when $t \rightarrow \infty$. Since the predictors is attached with this exponential term, when the second moment of the exponential term goes to zero, the predictors’ second moment will also go to zero. In this case, the variance of the predictors will decrease as time increases. Thus, the level of nonstationarity due to the predictors decreases.
            
        \subsection{Appendix 2 Tables}
            \begin{table}[H]
    			\begin{center}
    				\begin{tabular}{|c|c|c|c|c|c|} 
    					\hline
    					Variables & Mean & Std. Dev. & Skewness & Kurtosis & Auto \\ 
    					\hline
    					DY & 0.0310 & 0.0121 & 0.5313 & 2.5938 & 0.962 \\
    					\hline
    					EP & 0.0648 & 0.0261 & 0.9691 & 3.7927 & 0.959 \\
    					\hline
    					Bm & 0.5099 & 0.2485 & 0.6918 & 2.8151 & 0.977 \\
    					\hline
    					cay & $-2.63e^{-17}$ & 0.0226 & -0.2073 & 2.0706 & 0.959 \\
    					\hline
    					SR & 0.0459 & 0.0792 & -1.2549 & 5.8946 & 0.094\\
    					\hline
    				\end{tabular}
    				\caption{Summary Statistics for Quarterly S\&P500 Data}
    			\end{center}
    		\end{table}
    		
    		\begin{table}[H]
    			\begin{center}
    				\begin{tabular}{|c|c|c|} 
    					\hline
    					 & ADF test statistic & p-value \\ 
    					\hline
    					Dividend yield & -2.453278 & 0.1283 \\
    					\hline
    					Earnings-to-price ratio & -3.018963 & 0.0344 \\
    					\hline
    					Book-to-market ratio & -1.491306 & 0.5367 \\
    					\hline
    					Stock returns & -14.81049 & 0.0000 \\
    					\hline
    				\end{tabular}
    				\caption{ADF Test results for Dividend yield, Earnings-to-price ratio, Book-to-market ratio, and Stock returns using Quarterly data}
    			\end{center}
    		\end{table}
    		
    		\begin{table}[H]
    			\begin{center}
    				\begin{tabular}{|c|c|c|c|c|c|} 
    					\hline
    					Variables & Mean & Std. Dev. & Skewness & Kurtosis & Auto \\ 
    					\hline
    					DY & 0.0390 & 0.0177 & 0.9840 & 5.5043 & 0.981 \\
    					\hline
    					EP & 0.0711 & 0.0286 & 0.8378 & 3.2598 & 0.987 \\
    					\hline
    					Bm & 0.5588 & 0.2601 & 0.8299 & 4.6326 & 0.985 \\
    					\hline
    					SR & 0.0455 & 0.0177 & 1.0144 & 14.8548 & 0.157 \\
    					\hline
    				\end{tabular}
    				\caption{Summary Statistics for Monthly S\&P500 Data}
    			\end{center}
    		\end{table}
    		
    		\begin{table}[H]
    			\begin{center}
    				\begin{tabular}{|c|c|c|} 
    					\hline
    					 & ADF test statistic & p-value \\ 
    					\hline
    					Dividend yield & -3.148575 & 0.0234 \\
    					\hline
    					Earnings-to-price ratio & -3.226827 & 0.0187 \\
    					\hline
    					Book-to-market ratio & -3.089671 & 0.0276 \\
    					\hline
    					Stock returns & -12.30983 & 0.0000 \\
    					\hline
    				\end{tabular}
    				\caption{ADF Test results for Dividend yield, Earnings-to-price ratio, Book-to-market ratio, and Stock returns using Monthly data}
    			\end{center}
    		\end{table}
    		
    		\begin{table}[H]
    			\begin{center}
    				\begin{tabular}{|c|c|c|c|c|} 
    				    \hline
    				    Coefficients: & & & & \\
    					\hline
    					 & Estimate & Std. Error & t value & p-value \\ 
    					\hline
    					$y_{t-1}$ & 0.006233 & 0.061932 & 0.101 & 0.9199 \\
    					\hline
    					$y_{t-2}$ & -0.008595 & 0.062837 & -0.137 & 0.8913 \\
    					\hline
    					$y_{t-3}$ & -0.138523 & 0.062471 & -2.217 & 0.0275 \\
    					\hline
    					$y_{t-4}$ & -0.003129 & 0.062469 & -0.050 & 0.9601 \\
    					\hline
    					$x_{t-1, cay}$ & 0.166729 & 0.280236 & 0.595 & 0.5524 \\
    					\hline
    					$x_{t-1, DY}$ & 4.035963 & 3.995149 & 1.010 & 0.3134 \\
    					\hline
    					$x_{t-1, EP}$ & 0.040080 & 1.095308 & 0.037 & 0.9708 \\
    					\hline
    					$x_{t-1, Bm}$ & -0.031024 & 0.179398 & -0.173 & 0.8628 \\
    					\hline
    					\multicolumn{5}{|p{1.3cm}|}{\makecell{ Adjusted R-squared: 0.05601 \\ F-statistic: 2.429 (p-value: 0.006825) }} \\
    					\hline
    				\end{tabular}
    				\caption{Quarterly data in-sample estimation output}
    			\end{center}
    		\end{table}
    		
    		\begin{table}[H]
    			\begin{center}
    				\begin{tabular}{|c|c|c|c|c|} 
    				    \hline
    				    Coefficients: & & & & \\
    					\hline
    					 & Estimate & Std. Error & t value & p-value \\ 
    					\hline
    					$y_{t-1}$ & -0.040346 & 0.028792 & -1.401 & 0.161395 \\
    					\hline
    					$y_{t-2}$ & -0.079825 & 0.029043 & -2.749 & 0.006079 \\
    					\hline
    					$y_{t-3}$ & -0.005013 & 0.028945 & -0.173 & 0.862527 \\
    					\hline
    					$y_{t-4}$ & -0.043113 & 0.028769 & -1.499 & 0.134245 \\
    					\hline
    					$x_{t-1, DY}$ & 0.836885 & 0.242862 & 3.446 & 0.000589 \\
    					\hline
    					$x_{t-1, EP}$ & -0.119261 & 0.290809 & -0.410 & 0.681807 \\
    					\hline
    					$x_{t-1, Bm}$ & 0.008797 & 0.021860 & 0.402 & 0.687461 \\
    					\hline
    					\multicolumn{5}{|p{1.3cm}|}{\makecell{ Adjusted R-squared: 0.1236 \\ F-statistic: 17.55 (p-value $< 2.2e-16$) }} \\
    					\hline
    				\end{tabular}
    				\caption{Quarterly data in-sample estimation output}
    			\end{center}
    		\end{table}
    		
    		\begin{table}[H]
    			\begin{center}
    				\begin{tabular}{|c|c|c|c|c|} 
    					\hline
    					Models & RMSE $(j = 1)$ & RMSE $(j = 2)$ & RMSE $(j = 3)$ & RMSE $(j = 4)$ \\ 
    					\hline
    					Model 1-1 & 0.07440886 & 0.07381919 & 0.07491853 & 0.07298026 \\
    					\hline
    					Model 1-2 & 0.07892496 & 0.07624233 & 0.07174034 & 0.07174832 \\
    					\hline
    					Model 1-3 & 0.09336356 & 0.07540173 & 0.07280669 & 0.07394675 \\
    					\hline
    					Model 1-4 & 0.07213069 & 0.08321641 & 0.07772035 & 0.07451751 \\
    					\hline
    				\end{tabular}
    				\caption{RMSE output for multi-step $(j = 1, 2, 3, 4)$ ahead forecasts using quarterly data}
    			\end{center}
    		\end{table}
    		
    		\begin{table}[H]
    			\begin{center}
    				\begin{tabular}{|c|c|} 
    					\hline
    					Models & RMSE \\ 
    					\hline
    					Model 1-1 & 0.07403522 \\
    					\hline
    					Model 1-2 & 0.07472707 \\
    					\hline
    					Model 1-3 & 0.07932703 \\
    					\hline
    					Model 1-4 & 0.07700831 \\
    					\hline
    				\end{tabular}
    				\caption{RMSE output for overall forecasting performance using quarterly data}
    			\end{center}
    		\end{table}
    		
    		\begin{table}[H]
    			\begin{center}
    				\begin{tabular}{|c|c|c|c|c|} 
    					\hline
    					Models & RMSE $(j = 1)$ & RMSE $(j = 2)$ & RMSE $(j = 3)$ & RMSE $(j = 4)$ \\ 
    					\hline
    					Model 2-1 & 0.03868196 & 0.04350758 & 0.04258083 & 0.04288669 \\
    					\hline
    					Model 2-2 & 0.04490874 & 0.04479089 & 0.04464807 & 0.04469039 \\
    					\hline
    					Model 2-3 & 0.1163453 & 0.04364244 & 0.04453739 & 0.04386808 \\
    					\hline
    				\end{tabular}
    				\caption{RMSE output for multi-step $(j = 1, 2, 3, 4)$ ahead forecasts using monthly data}
    			\end{center}
    		\end{table}
    		
    		\begin{table}[H]
    			\begin{center}
    				\begin{tabular}{|c|c|c|c|c|} 
    					\hline
    					Models & RMSE $(j = 5)$ & RMSE $(j = 6)$ & RMSE $(j = 7)$ & RMSE $(j = 8)$ \\ 
    					\hline
    					Model 2-1 & 0.04322796 & 0.04313668 & 0.04252399 & 0.04269256 \\
    					\hline
    					Model 2-2 & 0.04491195 & 0.04501938 & 0.04463855 & 0.04462382 \\
    					\hline
    					Model 2-3 & 0.04367167 & 0.04381935 & 0.04367915 & 0.0438779 \\
    					\hline
    				\end{tabular}
    				\caption{RMSE output for multi-step $(j = 5, 6, 7, 8)$ ahead forecasts using monthly data}
    			\end{center}
    		\end{table}
    		
    		\begin{table}[H]
    			\begin{center}
    				\begin{tabular}{|c|c|c|c|c|} 
    					\hline
    					Models & RMSE $(j = 9)$ & RMSE $(j = 10)$ & RMSE $(j = 11)$ & RMSE $(j = 12)$ \\ 
    					\hline
    					Model 2-1 & 0.0428021 & 0.04294491 & 0.04259648 & 0.0418834 \\
    					\hline
    					Model 2-2 & 0.04403554 & 0.04399942 & 0.04401622 & 0.043272792 \\
    					\hline
    					Model 2-3 & 0.04343697 & 0.04415406 & 0.04444483 & 0.04378471 \\
    					\hline
    				\end{tabular}
    				\caption{RMSE output for multi-step $(j = 9, 10, 11, 12)$ ahead forecasts using monthly data}
    			\end{center}
    		\end{table}
    		
    		\begin{table}[H]
    			\begin{center}
    				\begin{tabular}{|c|c|} 
    					\hline
    					Models & RMSE \\ 
    					\hline
    					Model 2-1 & 0.04247248 \\
    					\hline
    					Model 2-2 & 0.04446577 \\
    					\hline
    					Model 2-3 & 0.05380378 \\
    					\hline
    				\end{tabular}
    				\caption{RMSE output for overall forecasting performance using monthly data}
    			\end{center}
    		\end{table}
        
        \section{Appendix 3 Graphs}
            \begin{figure}[H]
    			\centerline{\includegraphics[width=4.66in, height=3.33in]{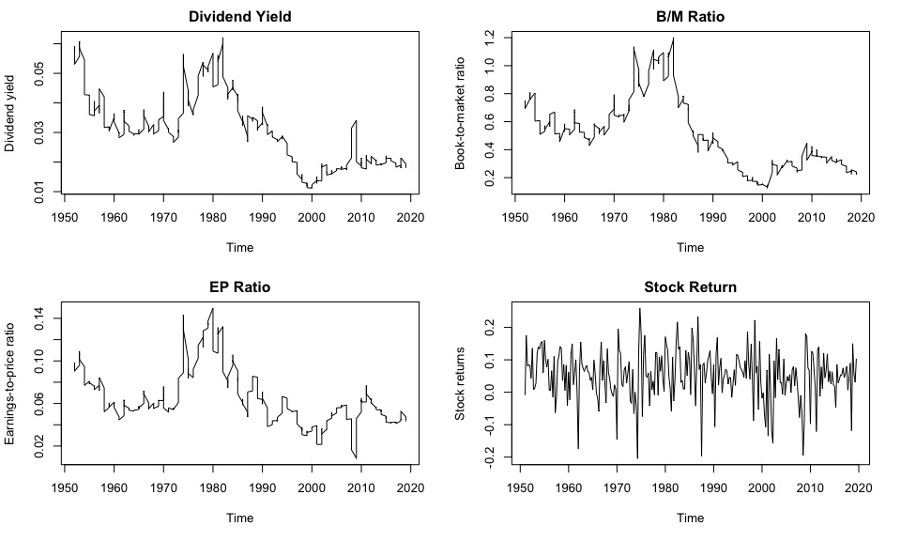}}
    				\caption{Time series plots of Dividend yield, Earnings-to-price ratio, Book-to-market ratio and Stock returns (Quarterly)}
    		\end{figure}
    		
    		\begin{figure}[H]
    			\centerline{\includegraphics[width=4.66in, height=3.33in]{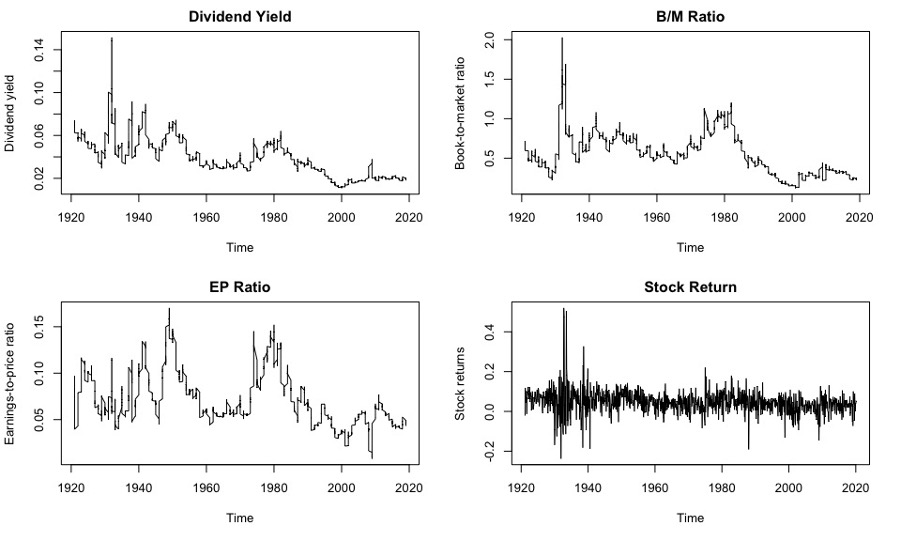}}
    				\caption{Time series plots of Dividend yield, Earnings-to-price ratio, Book-to-market ratio and Stock returns (Monthly)}
    		\end{figure}
    		
    		\begin{figure}[H]
    			\centerline{\includegraphics[width=4.66in, height=3.33in]{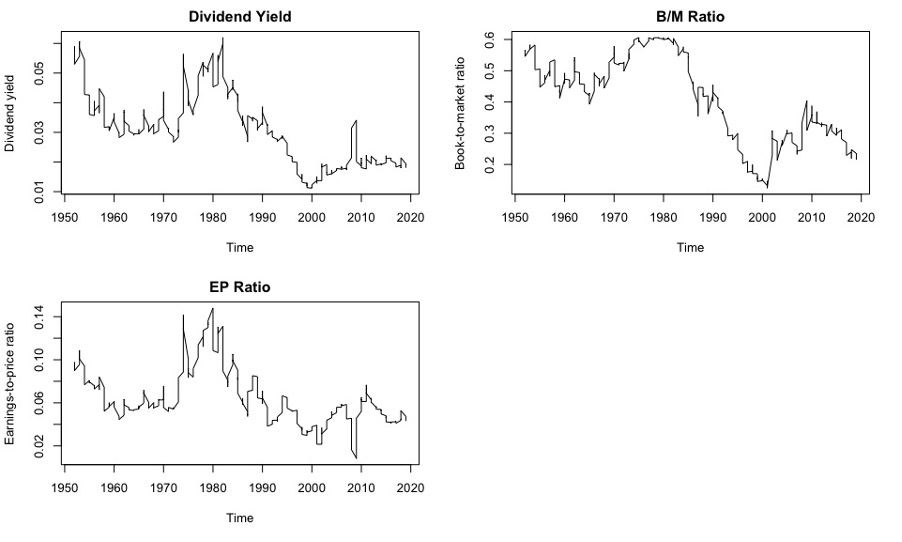}}
    				\caption{Time series plots of Dividend yield, Earnings-to-price ratio, Book-to-market ratio and Stock returns with exponential terms $(x_{t-1, k} \times e^{- \frac{x_{t-1, k}^2}{2}})$ (Quarterly)}
    		\end{figure}
    		
    		\begin{figure}[H]
    			\centerline{\includegraphics[width=4.66in, height=3.33in]{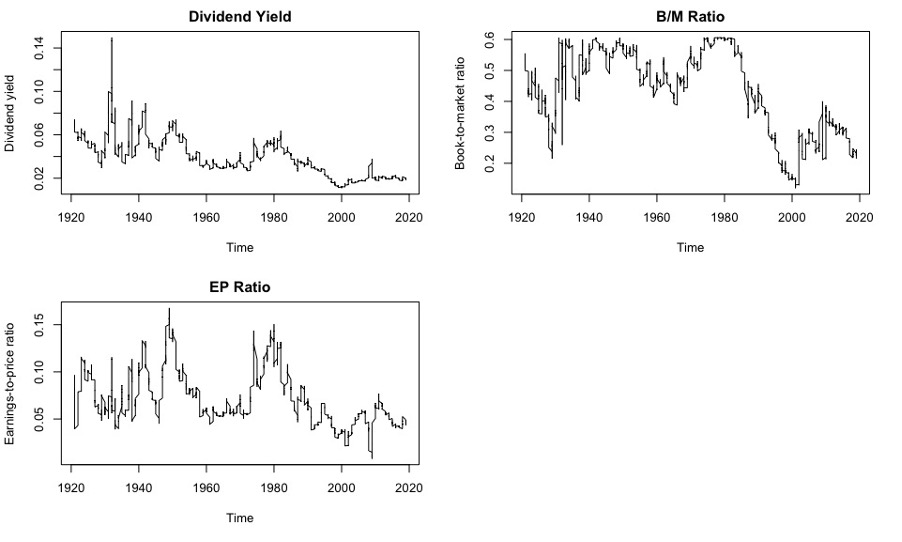}}
    				\caption{Time series plots of Dividend yield, Earnings-to-price ratio, Book-to-market ratio and Stock returns with exponential terms $(x_{t-1, k} \times e^{- \frac{x_{t-1, k}^2}{2}})$ (Monthly)}
    		\end{figure}
    		
    		\begin{figure}[H]
    			\centerline{\includegraphics[width=4.66in, height=3.33in]{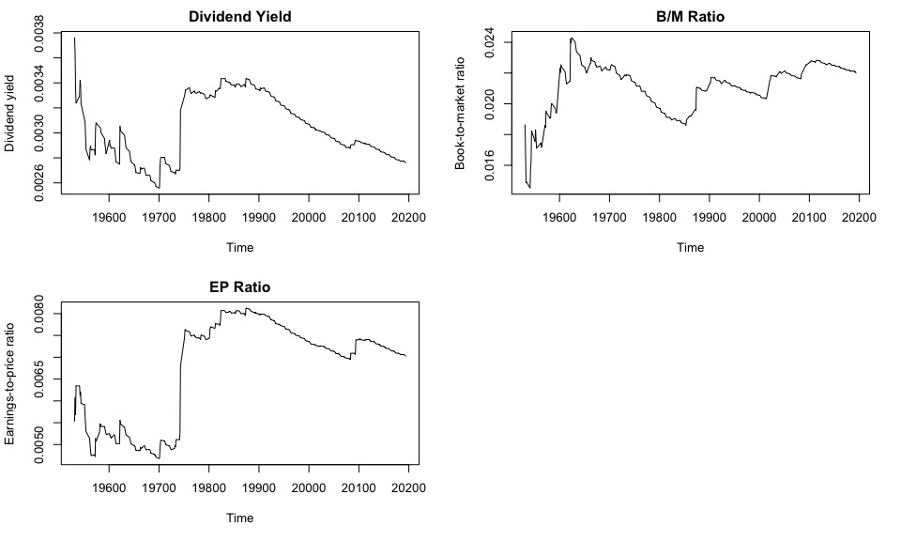}}
    				\caption{Standard Deviation of the First Derivative of the Time series plots of Dividend yield, Earnings-to-price ratio, Book-to-market ratio and Stock returns with exponential terms  $(x_{t-1, k} \times e^{- \frac{x_{t-1, k}^2}{2}})$ (Quarterly)}
    		\end{figure}
    		
    		\begin{figure}[H]
    			\centerline{\includegraphics[width=4.66in, height=3.33in]{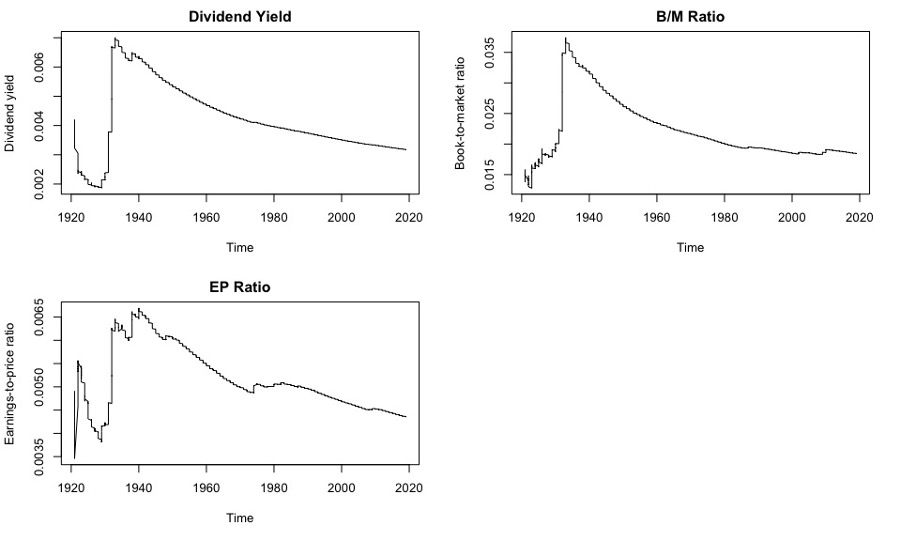}}
    				\caption{Standard Deviation of the First Derivative of the Time series plots of Dividend yield, Earnings-to-price ratio, Book-to-market ratio and Stock returns with exponential terms  $(x_{t-1, k} \times e^{- \frac{x_{t-1, k}^2}{2}})$ (Monthly)}
    		\end{figure}
    		
    		\begin{figure}[H]
    			\centerline{\includegraphics[width=4.66in, height=3.33in]{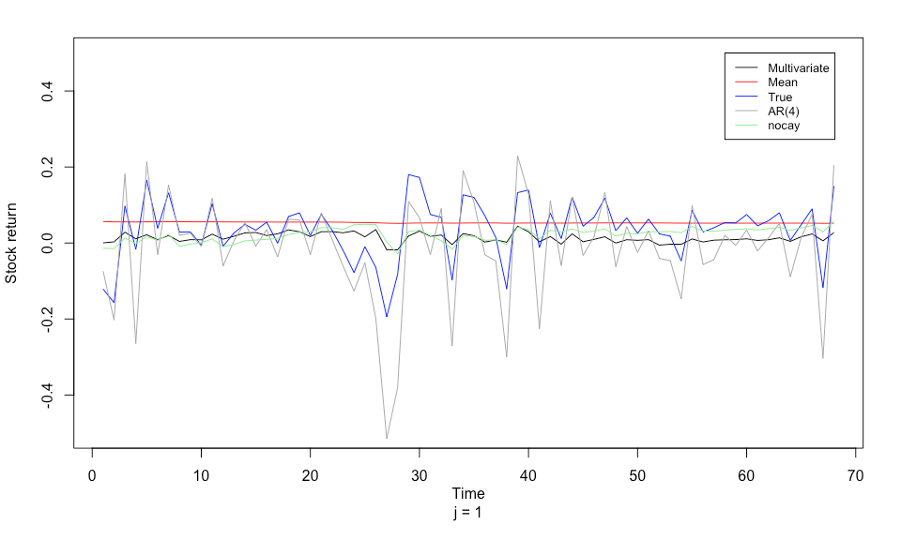}}
    				\caption{One-step ahead forecasts using quarterly data (Black line: Model 1-1; Red line: Model 1-2; Blue line: Real stock return; Grey line: Model 1-3; Green line: Model 1-4)}
    		\end{figure}
    		
    		\begin{figure}[H]
    			\centerline{\includegraphics[width=4.66in, height=3.33in]{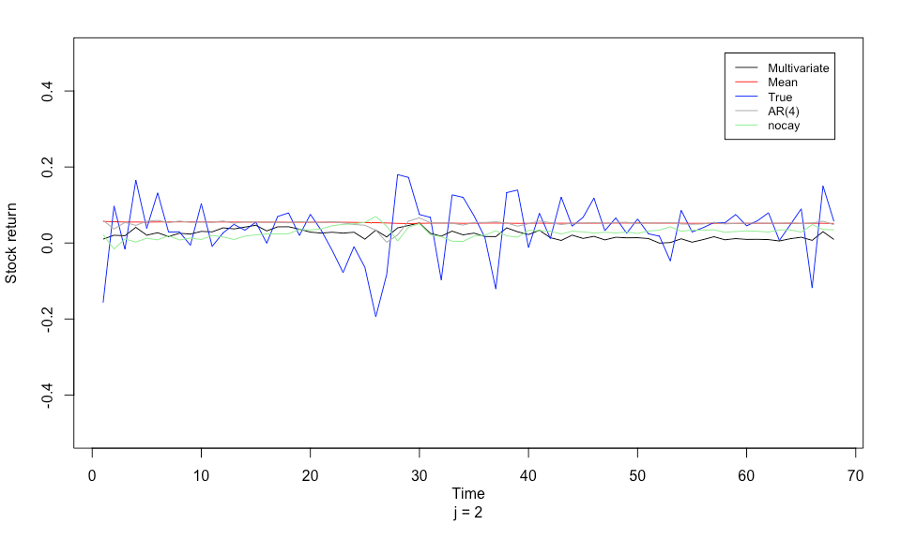}}
    				\caption{Two-step ahead forecasts using quarterly data(Black line: Model 1-1; Red line: Model 1-2; Blue line: Real stock return; Grey line: Model 1-3; Green line: Model 1-4)}
    		\end{figure}
    		
    		\begin{figure}[H]
    			\centerline{\includegraphics[width=4.66in, height=3.33in]{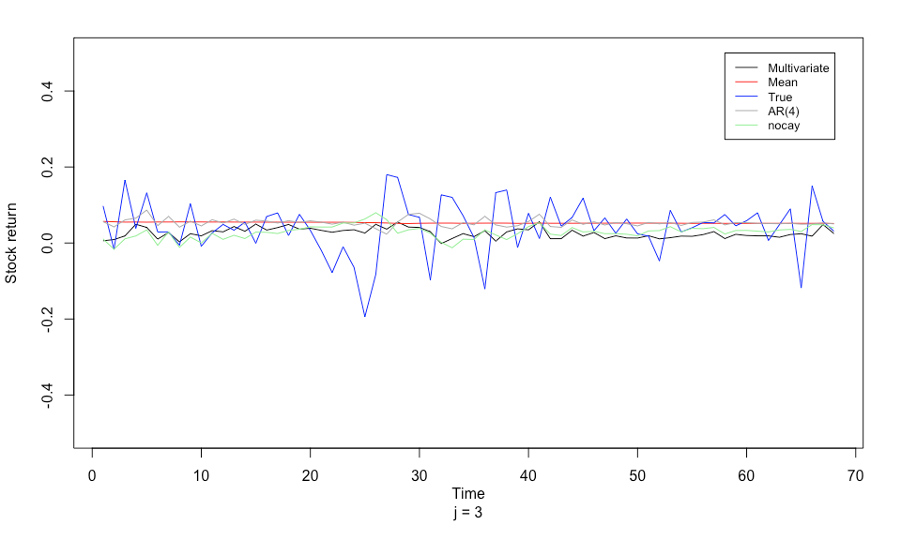}}
    				\caption{Three-step ahead forecasts using quarterly data(Black line: Model 1-1; Red line: Model 1-2; Blue line: Real stock return; Grey line: Model 1-3; Green line: Model 1-4)}
    		\end{figure}
    		
    		\begin{figure}[H]
    			\centerline{\includegraphics[width=4.66in, height=3.33in]{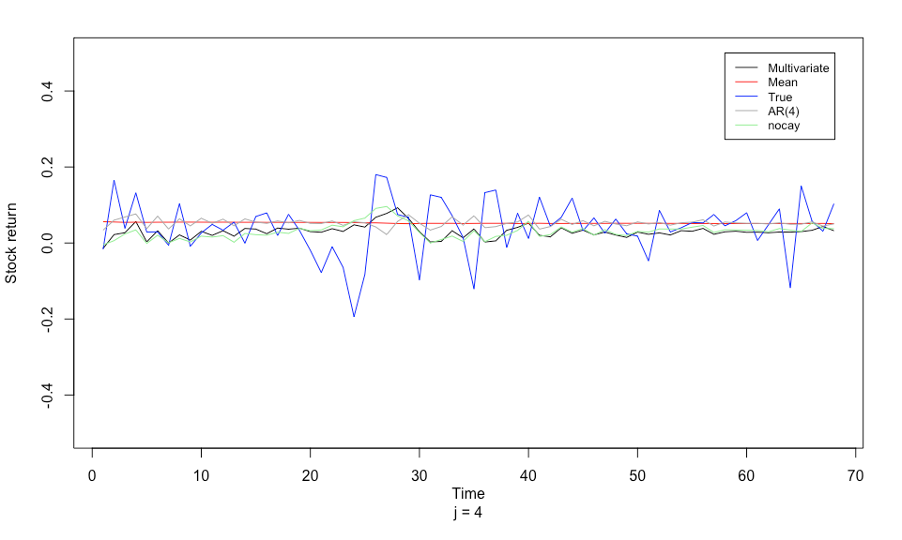}}
    				\caption{Four-step ahead forecasts using quarterly data(Black line: Model 1-1; Red line: Model 1-2; Blue line: Real stock return; Grey line: Model 1-3; Green line: Model 1-4)}
    		\end{figure}
    		
    		\begin{figure}[H]
    			\centerline{\includegraphics[width=4.66in, height=3.33in]{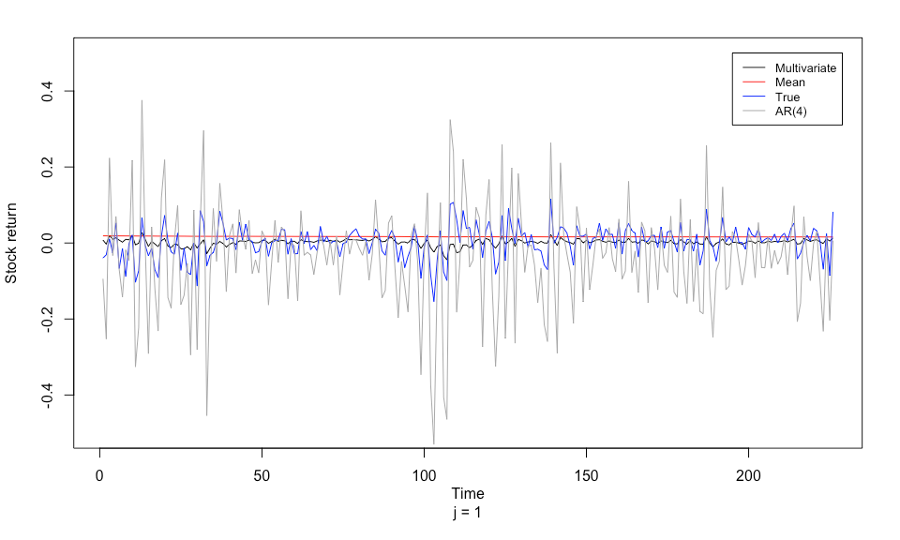}}
    				\caption{One-step ahead forecasts using monthly data(Black line: Model 2-1; Red line: Model 2-2; Blue line: Real stock return; Grey line: Model 2-3)}
    		\end{figure}
    		
    		\begin{figure}[H]
    			\centerline{\includegraphics[width=4.66in, height=3.33in]{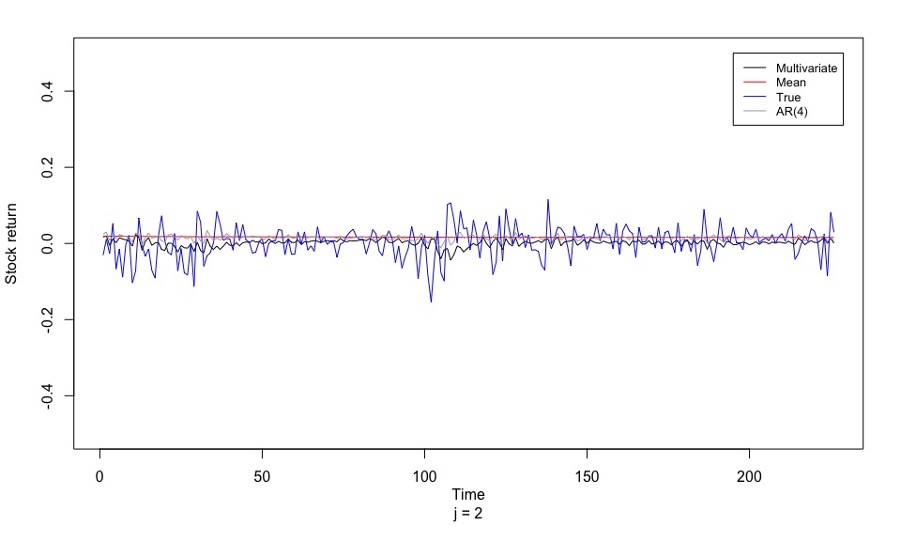}}
    				\caption{Two-step ahead forecasts using monthly data(Black line: Model 2-1; Red line: Model 2-2; Blue line: Real stock return; Grey line: Model 2-3)}
    		\end{figure}
    		
    		\begin{figure}[H]
    			\centerline{\includegraphics[width=4.66in, height=3.33in]{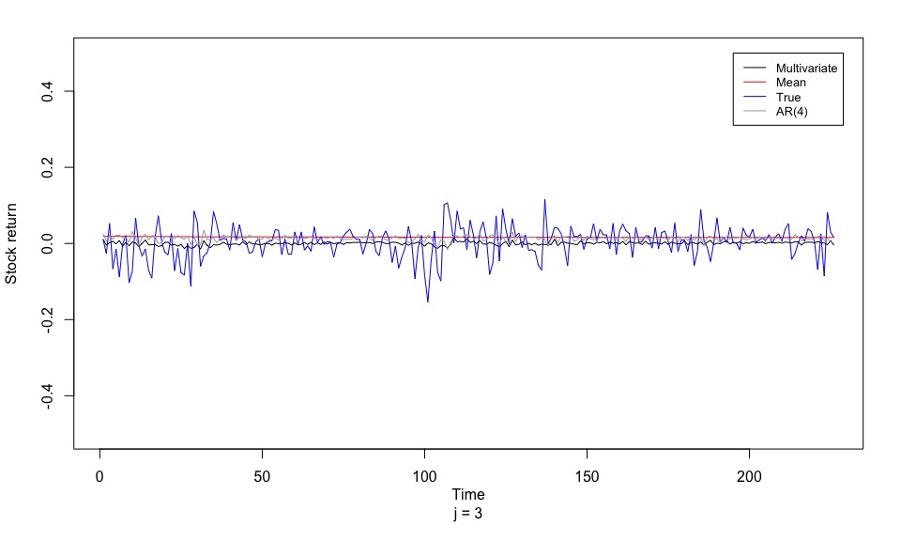}}
    				\caption{Three-step ahead forecasts using monthly data(Black line: Model 2-1; Red line: Model 2-2; Blue line: Real stock return; Grey line: Model 2-3)}
    		\end{figure}
    		
    		\begin{figure}[H]
    			\centerline{\includegraphics[width=4.66in, height=3.33in]{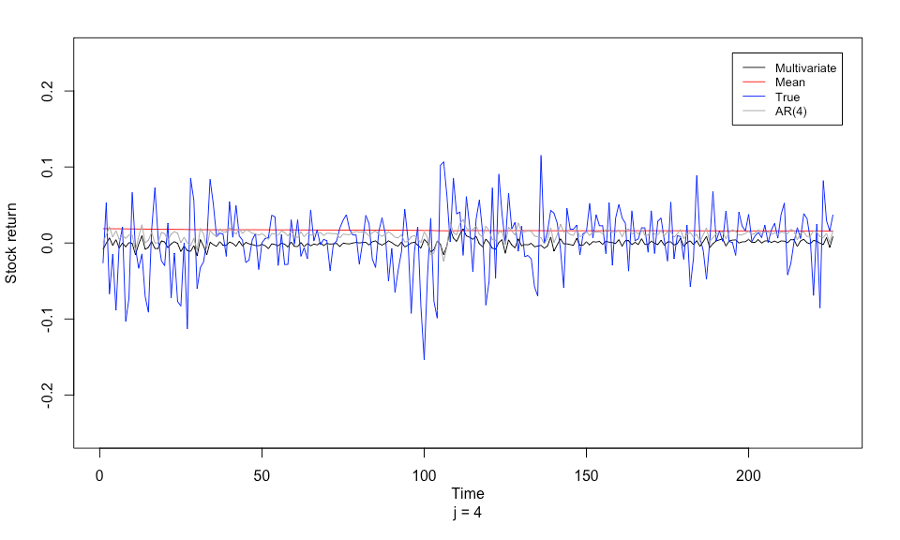}}
    				\caption{Four-step ahead forecasts using monthly data(Black line: Model 2-1; Red line: Model 2-2; Blue line: Real stock return; Grey line: Model 2-3)}
    		\end{figure}
    		
    		\begin{figure}[H]
    			\centerline{\includegraphics[width=4.66in, height=3.33in]{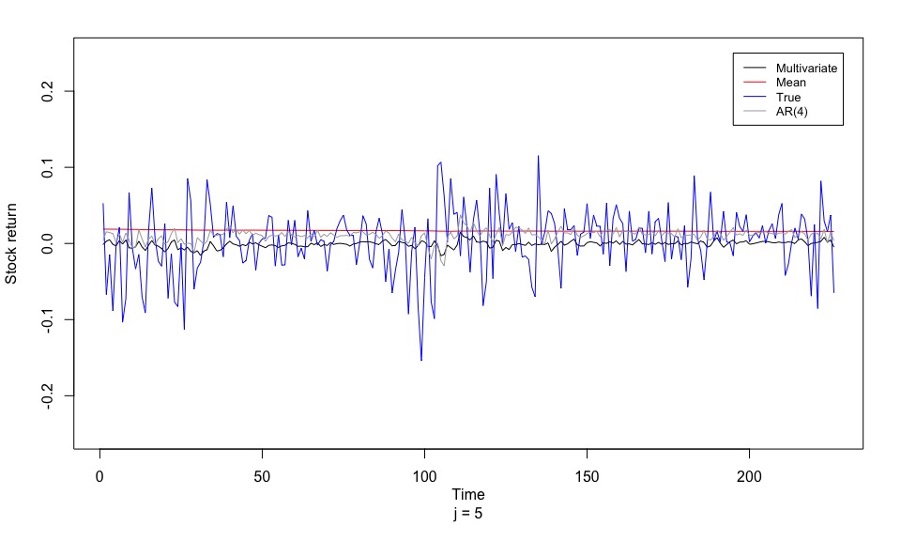}}
    				\caption{Five-step ahead forecasts using monthly data(Black line: Model 2-1; Red line: Model 2-2; Blue line: Real stock return; Grey line: Model 2-3)}
    		\end{figure}
    		
    		\begin{figure}[H]
    			\centerline{\includegraphics[width=4.66in, height=3.33in]{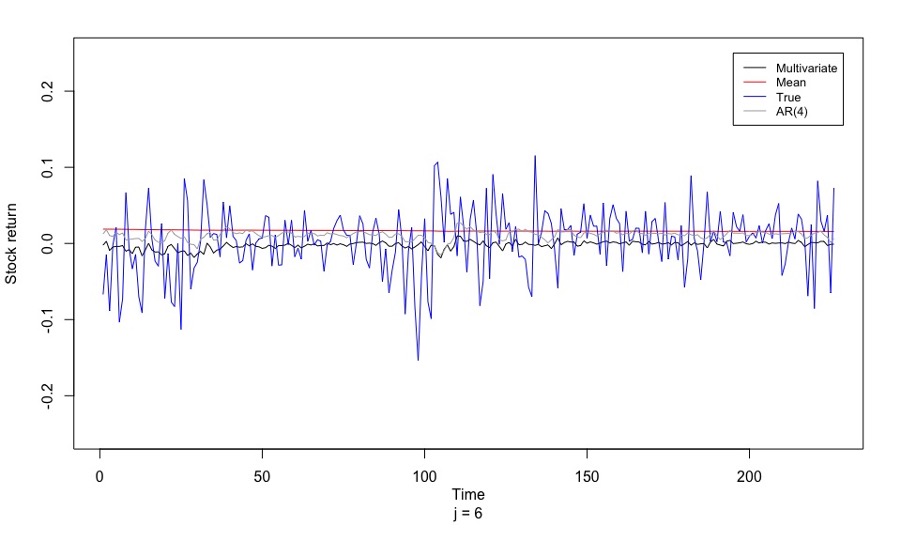}}
    				\caption{Six-step ahead forecasts using monthly data(Black line: Model 2-1; Red line: Model 2-2; Blue line: Real stock return; Grey line: Model 2-3)}
    		\end{figure}
    		
    		\begin{figure}[H]
    			\centerline{\includegraphics[width=4.66in, height=3.33in]{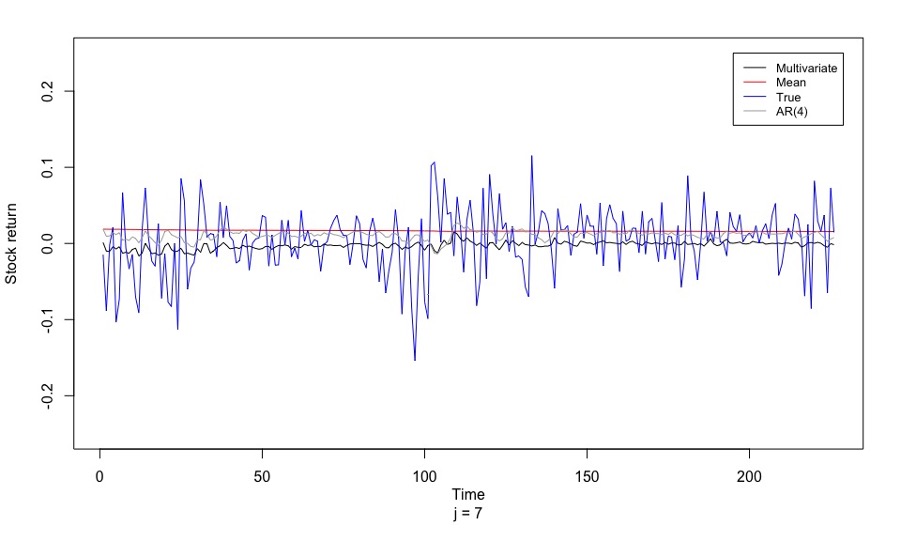}}
    				\caption{Seven-step ahead forecasts using monthly data(Black line: Model 2-1; Red line: Model 2-2; Blue line: Real stock return; Grey line: Model 2-3)}
    		\end{figure}
    		
    		\begin{figure}[H]
    			\centerline{\includegraphics[width=4.66in, height=3.33in]{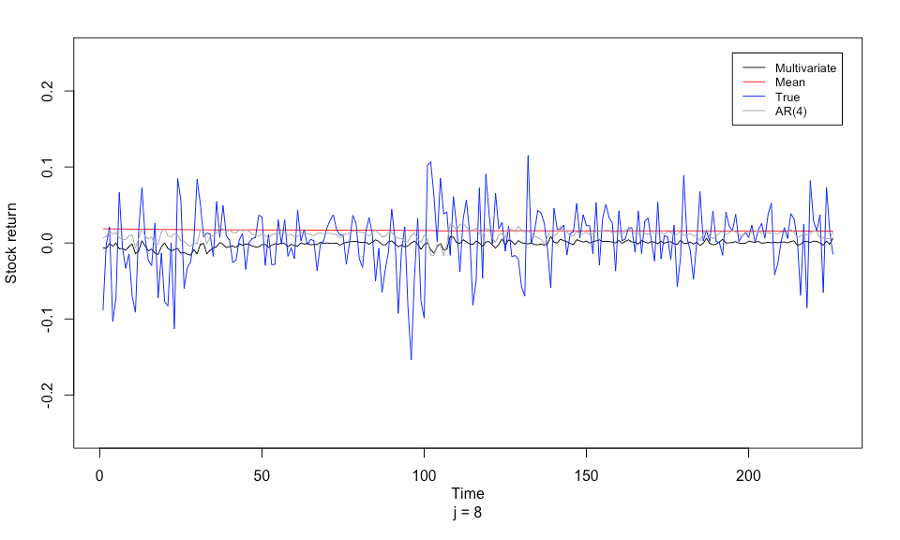}}
    				\caption{Eight-step ahead forecasts using monthly data(Black line: Model 2-1; Red line: Model 2-2; Blue line: Real stock return; Grey line: Model 2-3)}
    		\end{figure}
    		
    		\begin{figure}[H]
    			\centerline{\includegraphics[width=4.66in, height=3.33in]{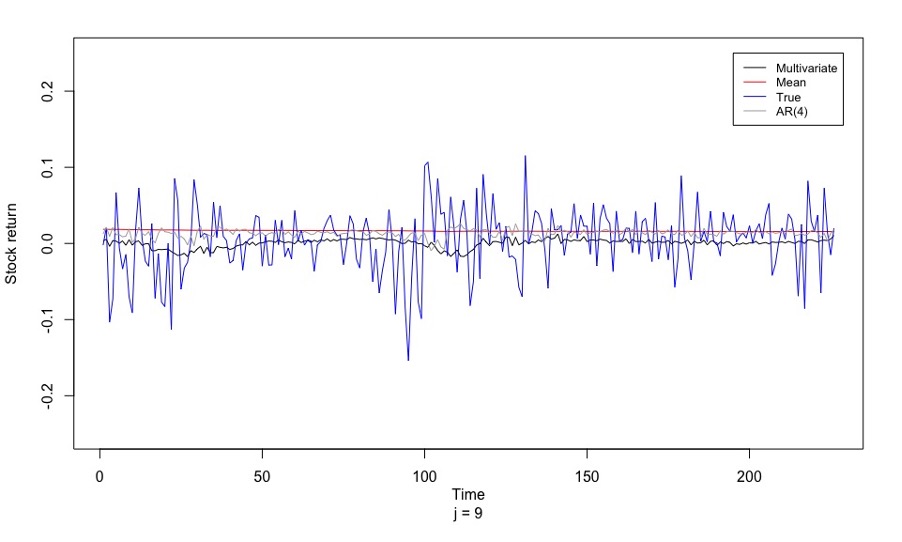}}
    				\caption{Nine-step ahead forecasts using monthly data(Black line: Model 2-1; Red line: Model 2-2; Blue line: Real stock return; Grey line: Model 2-3)}
    		\end{figure}
    		
    		\begin{figure}[H]
    			\centerline{\includegraphics[width=4.66in, height=3.33in]{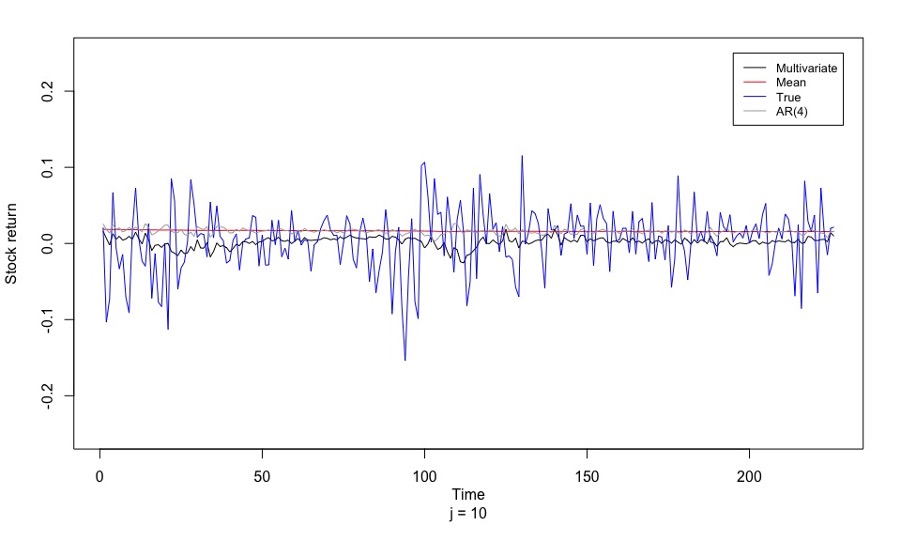}}
    				\caption{Ten-step ahead forecasts using monthly data(Black line: Model 2-1; Red line: Model 2-2; Blue line: Real stock return; Grey line: Model 2-3)}
    		\end{figure}
    		
    		\begin{figure}[H]
    			\centerline{\includegraphics[width=4.66in, height=3.33in]{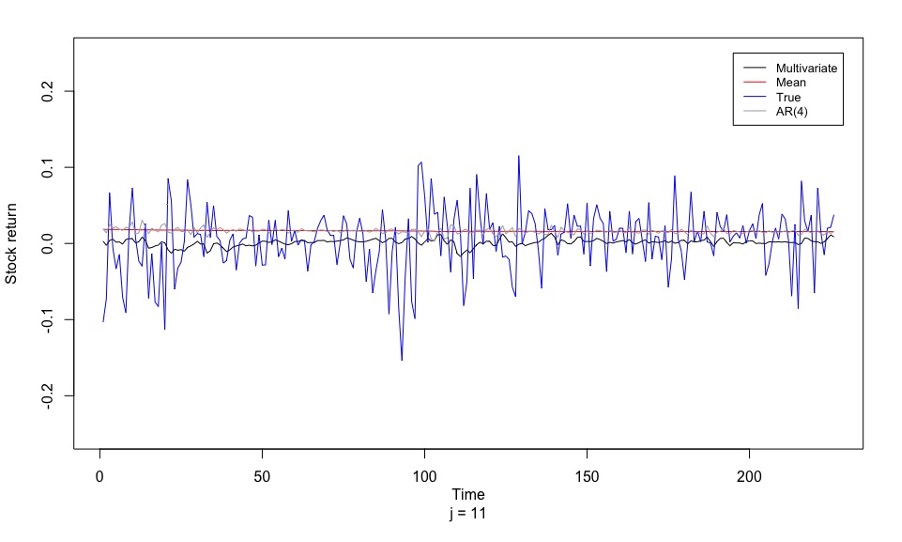}}
    				\caption{Eleven-step ahead forecasts using monthly data(Black line: Model 2-1; Red line: Model 2-2; Blue line: Real stock return; Grey line: Model 2-3)}
    		\end{figure}
    		
    		\begin{figure}[H]
    			\centerline{\includegraphics[width=4.66in, height=3.33in]{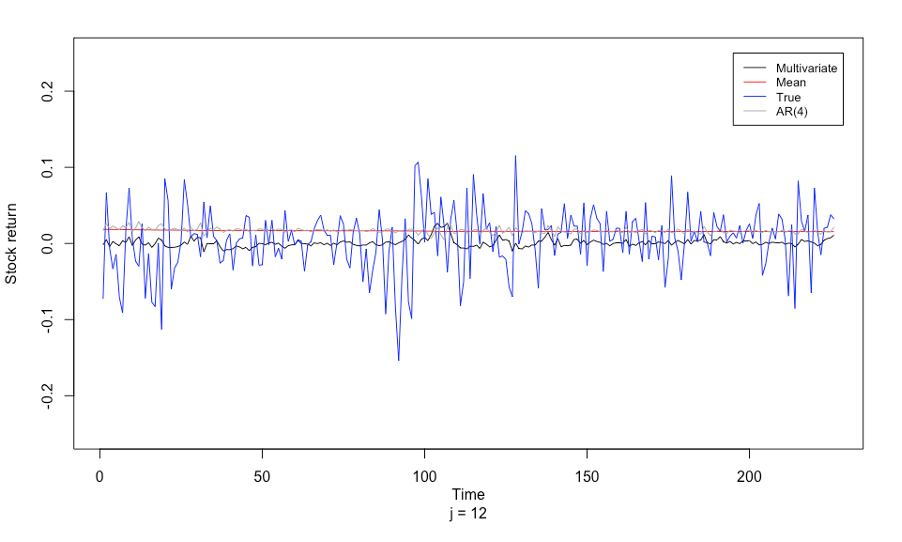}}
    				\caption{Twelve-step ahead forecasts using monthly data(Black line: Model 2-1; Red line: Model 2-2; Blue line: Real stock return; Grey line: Model 2-3)}
    		\end{figure}

\bibliography{mybibfile}

\begin{thebibliography}{10}
\expandafter\ifx\csname url\endcsname\relax
  \def\url#1{\texttt{#1}}\fi
\expandafter\ifx\csname urlprefix\endcsname\relax\def\urlprefix{URL }\fi
\expandafter\ifx\csname href\endcsname\relax
  \def\href#1#2{#2} \def\path#1{#1}\fi

\bibitem{fama1988permanent}
E.~F. Fama, K.~R. French, Permanent and temporary components of stock prices,
  Journal of political Economy 96~(2) (1988) 246--273.

\bibitem{fama1990stock}
E.~F. Fama, Stock returns, expected returns, and real activity, The journal of
  finance 45~(4) (1990) 1089--1108.

\bibitem{campbell2000asset}
J.~Y. Campbell, Asset pricing at the millennium, The Journal of Finance 55~(4)
  (2000) 1515--1567.

\bibitem{lettau2001consumption}
M.~Lettau, S.~Ludvigson, Consumption, aggregate wealth, and expected stock
  returns, the Journal of Finance 56~(3) (2001) 815--849.

\bibitem{campbell2006efficient}
J.~Y. Campbell, M.~Yogo, Efficient tests of stock return predictability,
  Journal of financial economics 81~(1) (2006) 27--60.

\bibitem{lettau2008reconciling}
M.~Lettau, S.~Van~Nieuwerburgh, Reconciling the return predictability evidence:
  The review of financial studies: Reconciling the return predictability
  evidence, The Review of Financial Studies 21~(4) (2008) 1607--1652.

\bibitem{welch2008comprehensive}
I.~Welch, A.~Goyal, A comprehensive look at the empirical performance of equity
  premium prediction, The Review of Financial Studies 21~(4) (2008) 1455--1508.

\bibitem{fama1989business}
E.~F. Fama, K.~R. French, Business conditions and expected returns on stocks
  and bonds, Journal of financial economics 25~(1) (1989) 23--49.

\bibitem{lamont1998earnings}
O.~Lamont, Earnings and expected returns, The journal of Finance 53~(5) (1998)
  1563--1587.

\bibitem{stambaugh1999predictive}
R.~F. Stambaugh, Predictive regressions, Journal of Financial Economics 54~(3)
  (1999) 375--421.

\bibitem{rozeff1984dividend}
M.~S. Rozeff, Dividend yields are equity risk premiums, Journal of Portfolio
  management (1984) 68--75.

\bibitem{eugene1992cross}
F.~Eugene, K.~French, The cross-section of expected stock returns, Journal of
  Finance 47~(2) (1992) 427--465.

\bibitem{kothari1997book}
S.~P. Kothari, J.~Shanken, Book-to-market, dividend yield, and expected market
  returns: A time-series analysis, Journal of financial economics 44~(2) (1997)
  169--203.

\bibitem{pontiff1998book}
J.~Pontiff, L.~D. Schall, Book-to-market ratios as predictors of market
  returns, Journal of financial economics 49~(2) (1998) 141--160.

\bibitem{nelson1993predictable}
C.~R. Nelson, M.~J. Kim, Predictable stock returns: The role of small sample
  bias, The Journal of Finance 48~(2) (1993) 641--661.

\bibitem{goyal2003predicting}
A.~Goyal, I.~Welch, Predicting the equity premium with dividend ratios,
  Management Science 49~(5) (2003) 639--654.

\bibitem{kostakis2015robust}
A.~Kostakis, T.~Magdalinos, M.~P. Stamatogiannis, Robust econometric inference
  for stock return predictability, The Review of Financial Studies 28~(5)
  (2015) 1506--1553.

\bibitem{ang2007stock}
A.~Ang, G.~Bekaert, Stock return predictability: Is it there?, The Review of
  Financial Studies 20~(3) (2007) 651--707.

\bibitem{cai2017simple}
B.~Cai, J.~Gao, A simple nonlinear predictive model for stock returns,
  Available at SSRN 3051006.

\bibitem{pesaran1995predictability}
M.~H. Pesaran, A.~Timmermann, Predictability of stock returns: Robustness and
  economic significance, The Journal of Finance 50~(4) (1995) 1201--1228.

\bibitem{engle1987co}
R.~F. Engle, C.~W. Granger, Co-integration and error correction:
  representation, estimation, and testing, Econometrica: journal of the
  Econometric Society (1987) 251--276.

\bibitem{lee1996comovements}
B.-S. Lee, Comovements of earnings, dividends, and stock prices, Journal of
  Empirical Finance 3~(4) (1996) 327--346.

\bibitem{duy1998modeling}
T.~A. Duy, M.~A. Thoma, Modeling and forecasting cointegrated variables: some
  practical experience, Journal of Economics and Business 50~(3) (1998)
  291--307.

\bibitem{inoue2004bagging}
A.~Inoue, L.~Kilian, Bagging time series models, Available at SSRN 540262.

\bibitem{wooldridge2015introductory}
J.~M. Wooldridge, Introductory econometrics: A modern approach, Cengage
  learning, 2015.

\bibitem{clark2009improving}
T.~E. Clark, M.~W. McCracken, Improving forecast accuracy by combining
  recursive and rolling forecasts, International Economic Review 50~(2) (2009)
  363--395.

\bibitem{hyndman2006another}
R.~J. Hyndman, A.~B. Koehler, Another look at measures of forecast accuracy,
  International journal of forecasting 22~(4) (2006) 679--688.

\bibitem{campbell2001valuation}
J.~Y. Campbell, R.~J. Shiller, Valuation ratios and the long-run stock market
  outlook: An update (2001).

\end{thebibliography}

\end{document}